\documentclass[12pt]{iopart}
\usepackage{iopams}
\usepackage{url}
\usepackage{graphicx,color}
\usepackage{bbm}

\expandafter\let\csname equation*\endcsname\relax
\expandafter\let\csname endequation*\endcsname\relax
\usepackage{amsmath}

\newcommand{\genotype}{\sigma} 
\newcommand{\moment}{\mu} 
\newcommand{\mlnx}{\chi} 
\newcommand{\bn}{\tau} 
\newcommand{\spin}{s} 
\newcommand{\R}{\mathbb{R}} 
\newcommand{\mutvec}{\vec{\xi}} 
\newcommand{\mutvecM}{\xi} 
\newcommand{\phevec}{\vec{z}} 
\newcommand{\fitftn}{f} 
\newcommand{\genfit}{W} 
\newcommand{\indic}{\mathfrak{I}} 
\newcommand{\Area}{{\cal A}} 
\newcommand{\AArea}{\widetilde {\cal A}} 
\newcommand{\Q}{\vec{Q}} 
\newcommand{\q}{\vec{q}} 
\newcommand{\vk}{\vec{k}} 
\newcommand{\D}{{\cal D}} 
\newcommand{\N}{\mathcal{N}} 
\newcommand{\wLi}{\Phi} 
\newcommand{\tsum}{\tilde a} 
\newcommand{\scaledN}{X} 
\newcommand{\G}{{\cal G}} 

\newcommand{\y}{\vec{y}}
\newcommand{\z}{\vec{z}}

\newcommand{\sor}{\mathrm{o}} 

\newcommand{\Avr}[1]{\left \langle #1 \right \rangle}

\begin{document}
\title{Distribution of the number of fitness maxima in Fisher's Geometric Model}
\author{Su-Chan Park$^1$, Sungmin Hwang$^2$, and Joachim Krug$^3$}
\address{$^1$ Department of Physics, The Catholic University of Korea, Bucheon 14662, Republic of Korea}
\address{$^2$ Capital Fund Management, 23-25 Rue de l'Universit\'e, 75007 Paris, France}
\address{$^3$ Institute for Biological Physics, University of Cologne,
  Z\"ulpicher Strasse 77, 50937 K\"oln, Germany}
\vspace{10pt}
\begin{indented}
\item[]\today
\end{indented}
\begin{abstract}
  Fisher's geometric model describes biological fitness landscapes by combining a linear map from the discrete space of genotypes to an $n$-dimensional
  Euclidean phenotype space with a nonlinear, single-peaked phenotype-fitness map. Genotypes are represented by binary sequences of length $L$, and
  the phenotypic effects of mutations at different sites are represented by $L$ random vectors drawn from an isotropic
  Gaussian distribution. Recent work has shown that the interplay between the genotypic and phenotypic levels gives rise to a range of different
  landscape topographies that can be characterised by the number of local fitness maxima. Extending our previous study of the mean number of local maxima,
  here we focus on the distribution of the number of maxima when the limit $L \to \infty$ is taken at finite $n$. We identify the typical scale
  of the number of maxima for general $n$, and determine the full scaled probability density and two point correlation function of maxima for
  the one-dimensional case. 
  We also elaborate on the close relation of the model to the anti-ferromagnetic Hopfield model with $n$ random continuous pattern vectors,
  and show that many
  of our results carry over to this setting. More generally, we expect that our analysis can help to elucidate the fluctuation structure of
  metastable states in various spin glass problems.
\end{abstract}
%
\submitto{\jpa}

\section{\label{Sec:intro}Introduction}
The concept of a fitness landscape has proven to be useful in
describing the dynamics of evolving biological populations~\cite{Orr2005,Visser2014,Fragata2019}. 
The fitness landscape is a mapping $W(\sigma)$ that assigns a fitness
value to each genetic sequence or \textit{genotype} $\sigma$~\cite{Visser2014,Fragata2019,Wright1931,Szendro2013a}. 
While natural selection can be conceptualised as a hill-climbing process
favouring fitter genotypes, random mutations generate and maintain the
genetic diversity that selection acts upon. 
Equipped with specific rules for the evolutionary dynamics, the
changes in genotype frequencies are given by transition rates $\sigma
\to \sigma'$ that typically depend on the fitness differences
$W(\sigma') - W(\sigma)$ between neighbouring genotypes \cite{Gillespie1984,Orr2002,Sella2005}.
Thus, determining the functional form of $ W(\sigma)$ is a crucial step for modelling the evolution of populations.

Instead of considering a single instance of a fitness function, one often defines random fitness landscape ensembles 
	based on a plausible set of assumptions \cite{Szendro2013a,Stadler1999,Hwang2018}. 
By studying the statistical properties of such ensembles,
topographical features of typical fitness landscapes corresponding to
a given set of assumptions can be inferred.
One large class of fitness landscape ensembles are \emph{phenotypic fitness landscapes}.
These ensembles introduce an intermediate phenotypic space
\cite{Domingo2019,Manrubia2020} that
mediates the mapping from genotype to fitness through a relation of
the form $W(\sigma) = f(\vec{z}(\sigma))$, where $\vec{z}(\sigma)$ is
the phenotype and $f(\vec{z})$ the phenotype-fitness map.  

Fisher's geometric model (FGM) is the paradigmatic representative of a
phenotypic fitness landscape ensemble~\cite{Fisher1930,Martin2007,Gros2009,Martin2014,Blanquart2014,Tenaillon2014,Hwang2017}.
Apart from additional model-specific settings, it shares three major ingredients:
i) An organism is characterised by a phenotype represented by a vector $\vec{z} = (z_1, z_2,
\cdots, z_n)$ in an $n$-dimensional Euclidean space. The real-valued
components $z_i$ describe quantitative traits of the organism such as its body mass or size.  
ii) Mutations in the genotype space induce random displacements $\mutvec$ of phenotypes, by which the population explores the phenotype space.
Importantly, the random displacements corresponding to different
mutations are added vectorially \cite{Martin2007}. 
iii) A single-peaked fitness function $ f(\vec{z})$ forms
nonlinear fitness isoclines by which genotype-genotype
interactions emerge \cite{Blanquart2014,Schoustra2016}. The peak of
$f(\vec{z})$ defines the location of the optimal phenotype which can
be placed at the origin $\vec{z} = 0$ of the trait space without loss
of generality.

Having identified these elements, it is not difficult to establish a connection between FGM and disordered discrete spin models.
It is based on three observations. First, the presence or absence
of a mutation is encoded by a binary variable $\tau_i = \{0,1\}, i = 1,\cdots,L$, which
can alternatively be represented by an Ising spin $s_i = \{-1,
1\}$. Second, the fitness plays the role of a Hamiltonian of the form $
-f(\vec{z}(\sigma))$ and third, the fitness function is determined by
the choice of the random displacements $\mutvec$, which introduce
quenched disorder into the problem.

In fact, it will be shown below that FGM shares a close similarity with the celebrated Hopfield model of associative memory~\cite{Hopfield1982,Amit1989,Hertz1991}.
In this model, the Hamiltonian is designed such that a set of
predefined patterns
are the attractors of the corresponding dynamics.
Thus, if the initial configuration is closest to one of the stored
patterns, it can find it through the dynamics as long as the system is in the retrieval phase.
These patterns correspond to the random mutational displacement
vectors in FGM, but the interactions turn out to be antiferromagnetic, 
in the sense that the spin configurations try to avoid predefined
patterns.
The antiferromagnetic Hopfield model \cite{Nokura1998} (AFHM) has been studied in various contexts such as the random
orthogonal model~\cite{Cherrier2003} or minority
games~\cite{Challet1997,Marsili2000,Challet2000,Chakraborti2015}.
In the present work we will be particularly concerned with the
one-dimensional AFHM which is closely related to the number partitioning
problem \cite{Ferreira1998,Mertens2000}.  

In our recent contribution~\cite{Hwang2017}, we have performed a
detailed analysis of the mean number of local maxima in FGM and
determined the phase diagram of the model. Three distinct phases were
identified which correspond to different mechanisms by which
genotype-genotype interactions and multiple fitness peaks are created.
These results were however mostly limited to the mean number of local
maxima despite our observation that the number of maxima fluctuates strongly
in the limit $L \to \infty$.
Here, we address this issue by computing the higher order moments of
the number of maxima. In the case of a one-dimensional phenotype space
($n=1$) this enables us to determine the full distribution of the
number of maxima, which turns out to have a highly nontrivial shape. 

In the context of disordered spin systems, the question addressed in
this article can be phrased differently: How many metastable states that are stable under single spin flips\footnote{Note that these states do not necessarily correspond to metastable phases in the thermodynamic sense.}
exist at zero temperature?
This type of question has been studied in various spin glass models using the so-called Tanaka-Edwards formalism~\cite{Cherrier2003,Ferreira1998,Tanaka1980,Bray1980,Bray1981,Gardner1986,Treves1988,Singh1995}. 
Despite these similarities, an advantage of our model in terms
of solvability compared to other spin models relies on the fact that the quantities of interest can be written geometrically. 
As will be shown below, this feature provides a powerful tool for studying
higher-order statistics.

The precise mathematical definition of FGM is provided in the next section. We then
discuss the relation to spin models and establish the approximate
equivalence with the AFHM in a scaling limit. The calculation of the
moments of the number of fitness maxima for general $n$ is explained
in \sref{Sec:moments}, and in \sref{Sec:exact} we specialise to the one-dimensional
case. In \sref{Sec:correlations} we derive the pair correlation function of maxima for
the one-dimensional model, and conclude in \sref{Sec:summary} with a summary and
a discussion of the broader context of our work. Detailed derivations
are mostly relegated to the appendices. 

\section{Fisher's geometric model}
Following a common convention in population genetics, a genotype is represented by a binary sequence of length $L$.
We denote such a sequence by $\genotype$, which sometimes carries an index like $\genotype_\alpha$.
The binary number appearing at the $i$th site of the sequence $\genotype$ 
is denoted by $\bn_i(\genotype)$ or simply by $\bn_i$ if the genotype under consideration is clear from the context.
The sequence with $\bn_i= 0$ for all $i$ will be called the wild-type genotype.
In biological terms, $\bn_i$ represents the presence ($\bn_i = 1$) or
absence ($\bn_i = 0$) of a
mutation at site $i$ with respect to the wild type.

A phenotype is represented by a vector in the $n$-dimensional
Euclidean trait space $\R^n$. As a consequence of the assumption
of additivity of mutational effects on the phenotype \cite{Martin2007},  
the phenotype vector $\phevec$ corresponding to a genotype $\genotype$ is constructed as
\begin{equation}
	\phevec(\genotype) = \Q + \sum_{i=1}^L \bn_i \mutvec_i,
\label{Eq:gen_phe_map}
\end{equation}
where $\Q$ is the wild-type phenotype and $\mutvec_i$ describes the
change in the phenotype due to a point mutation at site $i$.
The $\mutvec_i$'s are taken to be independent and identically distributed (i.i.d.)
random vectors drawn from a common probability density $p(\mutvec)$.
For convenience we usually choose $p(\vec{\xi})$ as a multivariate Gaussian distribution
\begin{equation}
p(\mutvec) \equiv \frac{1}{(2\pi)^{n/2}} \exp\left ( -\frac{1}{2} |\mutvec|^2 \right ),
	\label{Eq:common_dist}
\end{equation}
but most of our results readily generalise to other probability
densities that have a finite variance and non-vanishing weight at the
origin. In \eqref{Eq:common_dist} the variance has been set to unity,
which implies that distances in the
trait space are measured in units of single mutational effects. In
particular, $\vert \Q \vert$ is proportional to the minimal number of
mutations required to reach the fitness optimum from the wild type. 


By composing \eref{Eq:gen_phe_map} with a phenotype-fitness map
$f(\phevec)$, we obtain the $L$-dimensional genotypic fitness landscape
\begin{equation}
  \genfit(\genotype) \equiv \fitftn(\phevec(\genotype)).
\end{equation}
In the class of models know as FGM the phenotype-fitness function is
taken to be single peaked, with the unique phenotypic optimum
located at $\phevec = 0$. We will also assume isotropy in trait space,
which implies that $f$ depends only on $\vert \phevec
\vert$. Different choices for the shape of the fitness peak have been
considered in the literature \cite{Gros2009}, and statistical analyses have been
employed to infer the shape function, the dimensionality of trait
space, $n$, and the distance of the wild type to the peak, $\vert \Q
\vert$, from experimental data \cite{Schoustra2016,Blanquart2016,Weinreich2013}. 

In this paper, we are interested in how the number $\N$ of local fitness maxima 
in the genotypic landscape $W(\sigma)$  is distributed for large $L$. 
Here, by a local maximum we mean a genotype whose fitness is larger
than that of all $L$ neighours that can be reached by adding ($\tau_i
= 0 \to 1$) or removing ($\tau_i = 1 \to 0$) a single mutation. 
Since fitness is a decreasing function of
the magnitude of the phenotype vector $\phevec$, the condition that a genotype is a local
maximum is purely determined by the ordering of $|\phevec|$. 
Thus, we do not need specify the precise form of the phenotype-fitness
map $\fitftn(\phevec)$ for our purposes.

\section{\label{Sec:HM}Comparison to the antiferromagnetic Hopfield
  model}

\subsection{FGM as a spin model}
Our problem is identical to counting the number 
of local \textit{minima} of the quadratic Hamiltonian defined as
\begin{equation}
  \label{H_FGM}
H_\mathrm{FGM} \equiv  |\phevec(\genotype) |^2
=  |\Q + \sum_i \bn_i \mutvec_i |^2
=  |\Q|^2 + \sum_{ij}\mutvec_i\cdot \mutvec_j \bn_i \bn_j
+ 2 \Q \cdot \sum_i \bn_i \mutvec_i.
\end{equation}
Minimizing the last (linear) term simply amounts to setting $\tau_i =
1$ ($\tau_i = 0$) whenever $\Q \cdot \mutvec_i < 0$ ($\Q \cdot
\mutvec_i > 0$). For large $\vert \Q \vert$ this term dominates and
the fitness landscape becomes approximately additive
\cite{Hwang2017}. 

To elucidate the meaning of the quadratic term we set
$\Q = 0$ and rewrite \eqref{H_FGM} in terms of the Ising spins $s_i \equiv
2 \tau_i -1$. This yields 
\begin{equation}
  \label{H_FGM_spin}
H_\mathrm{FGM}^{\Q = 0} = \frac{1}{4} \sum_{ij} J_{ij} \spin_i \spin_j 
+ \frac{1}{2} \sum_i \tilde{h}_i \spin_i
+ \frac{1}{4} \sum_{ij} J_{ij},
\end{equation}
where
\begin{equation} 
J_{ij} \equiv \sum_{k=1}^n \mutvecM_i^k \mutvecM_j^k, \;\;\;\; 
\tilde{h}_i =  \sum_{k=1}^n  \left ( \sum_{j=1}^L \mutvecM_j^k \right ) \mutvecM_i^k,
\label{Eq:Ham_spin}
\end{equation}
and $\mutvecM_i^k$ is the $k$th component of $\mutvec_i$.
Since the last term in the Hamiltonian is a global constant for
a given realization of $\mutvec_i$'s, we can remove it without affecting the structure
of the energy landscape. Up to a conventional scale factor $\frac{1}{L}$, the
interaction term in \eqref{H_FGM_spin} is identical to the Hamiltonian
of the antiferromagnetic Hopfield model
\begin{equation}
  \label{Eq:H_AFM}
  H_\mathrm{AFHM} = \frac{1}{4L} \sum_{ij} J_{ij} s_i s_j =
  \frac{1}{4L} \sum_{k=1}^n \left( \sum_{i=1}^L \mutvecM_i^k s_i
  \right)^2
\end{equation}
with $n$ real-valued pattern vectors $(\mutvecM_1^k,
\mutvecM_2^k,\dots, \mutvecM_L^k) \in \mathbb{R}^L$. The AFHM
Hamiltonian is minimised by spin configurations that are maximally
orthogonal to the patterns \cite{Nokura1998}.

The FGM Hamiltonian differs from the AFHM by the presence of the random fields
$\tilde{h}_i$ which are determined by the pattern vectors through \eqref{Eq:Ham_spin}. As a consequence the fields are correlated with
the couplings $J_{ij}$. Although we will argue in the next subsection
that these correlated random fields become negligible at least in certain limits,
they enforce two fundamental differences between the
two models. First, the random fields break the
$s_i \to -s_i$ Ising symmetry of $H_\mathrm{AFHM}$. This symmetry implies in
particular that the number of local energy minima $\N$ has to be even
for the AFHM, while no such constraint applies for FGM. Second, the
correlations between the fields and the couplings ensure that the
ground state value $H_\mathrm{FGM}^{\Q = 0} = 0$ is 
realised by $s_i \equiv -1$ ($\tau_i = 0$), as is evident from the
construction of the model. By contrast, the ground state of
$H_\mathrm{AFHM}$ is nontrivial and generally unknown. 

\subsection{Joint limit $L, n \to \infty$}
Under the Gaussian distribution \eqref{Eq:common_dist} for the
displacement vectors the interior sum in the definition of the $\tilde{h}_i$
in \eqref{Eq:Ham_spin} can be written as  
\begin{equation}
\sum_{j=1}^L \mutvecM_j^k   = \sqrt{L} \eta_k,
\end{equation}
where the $\eta_k$'s are i.i.d.~Gaussian random variables with unit variance.
Moreover, since
\begin{equation}
	\left \langle \eta_k \mutvecM_i^k \right \rangle =\frac{1}{\sqrt{L}} \left \langle 
	\left ( \mutvecM_i^k\right )^2\right \rangle = \frac{1}{\sqrt{L}},\quad
	\left \langle \left ( \eta_k \mutvecM_i^k \right )^2
	\right \rangle 
	= 1 + \Or(L^{-1}),
\end{equation}
we can apply the central limit theorem to obtain
\begin{equation}
	\tilde{h}_i = \sqrt{L} \sum_{k=1}^n \eta_k \xi_i^k \approx \sqrt{Ln} h_i,
\end{equation}
where the $h_i$ are i.i.d.~Gaussian random variables with zero mean
and unit variance that become approximately independent of the 
$\xi_j^k$ in the joint limit $L, n\rightarrow \infty$. 

Specifically, if we take the limit $L \to \infty$ with $\alpha = n / L$ fixed,  
the FGM Hamiltonian formally maps to the AFHM with random fields
of strength $\sqrt{\alpha}$,
\begin{equation}
\frac{1}{L} H_\mathrm{FGM}^{\Q = 0} \approx \frac{1}{4L} \sum_{ij} J_{ij} \spin_i \spin_j 
+  \frac{1}{2} \sqrt{\alpha} \sum_i h_i \spin_i .
\label{Eq:AFH}
\end{equation}
The correlations between the couplings $J_{ij}$ and the
random fields $h_i$ in \eqref{Eq:AFH} can be estimated using Wick's
theorem, which yields
\begin{equation}
	\frac{1}{L}\left \langle J_{ij} h_p \right \rangle
	\approx \frac{1}{L\sqrt{Ln}}\left \langle \sum_k 
	\mutvecM_i^k
	\mutvecM_j^k
	\sum_{l=1}^n \sum_{q=1}^L 
	\mutvecM_p^l
	\mutvecM_q^l
      \right \rangle \sim
      \begin{cases}
        \Or(\alpha) & i=j \\
        \Or\left(\frac{1}{L}\right) & i=p \;\; \mathrm{or} \;\;
        j = p.
      \end{cases}
\end{equation}
This suggests that FGM and the AFHM without random fields should behave similarly at least
when $\alpha$ is small. A precise comparison can be made on the level
of  the exponential growth rate of the expected number of fitness peaks
$\langle \N\rangle$ defined by \cite{Hwang2017}
\begin{equation}
	\Sigma^* = \lim_{L\rightarrow \infty} \frac{ \ln \langle \N \rangle}{L}.
	\label{Eq:Sigma_star}
      \end{equation}
In \ref{App:Q0} we compute $\Sigma^\ast$ for FGM, which behaves as  
\begin{equation}
  \label{Eq:SigmaFGM}
	\Sigma^*_\text{FGM} \simeq \ln 2 - \frac{\alpha}{2}  \ln \left
          ( -\frac{4\ln \alpha}{\rme \alpha}\right ) = \ln 2 -\frac{\alpha}{2}
        \ln\left(\frac{\vert \ln \alpha \vert}{\alpha}\right) - \frac{\alpha}{2} \ln\left(\frac{4}{\rme}\right) 
      \end{equation}
      for $\alpha \to 0$. This should be compared with the result for
      the AFHM without random fields given by~\cite{Cherrier2003}\footnote{In \cite{Cherrier2003}, $\alpha'$ is used in place of $\alpha$.}
\begin{equation}
	\Sigma^*_\text{AFH} \simeq \ln 2 - \frac{\alpha}{2}
	\ln \left ( -\frac{2\ln \alpha}{\rme \alpha}\right ).
      \end{equation}
      The two expressions are seen to agree in the leading nontrivial
      behaviour, which shows that the correlated random fields in
      \eref{Eq:AFH} contribute only at the subleading order $\Or(\alpha)$.

      In the following sections we focus on the case of finite $n$, with
particular emphasis on the one-dimensional model.

\section{\label{Sec:moments}Moments} 
The number $\N$ of local fitness maxima
in the genotypic landscape can be formally written as
\begin{equation}
\N = \sum_{\genotype} \indic(\genotype), 
\end{equation}
where $\indic(\genotype)$ is an indicator that takes the value
1 if $\genotype$ is a local maximum and 0 otherwise.
We begin by writing a formal expression for the $m$th moment
\begin{equation}
\langle \N^m  \rangle
= \sum_{\genotype_1,\ldots,\genotype_m} 
\left \langle \prod_{\alpha=1}^m \indic\left ({\genotype_\alpha} \right ) \right \rangle,
\end{equation} 
where $\langle \cdots \rangle$ stands for  the average
over the ensemble of $\mutvec_i$'s.
Since $\langle \indic({\genotype_1}) \indic({\genotype_2})\ldots \indic({\genotype_m}) \rangle$ is simply the 
joint probability $P_m(\genotype_1,\genotype_2,\ldots,\genotype_m)$ 
that the indicated genotypes are local maxima,
we can rewrite the $m$th moment as
\begin{equation}
\langle \N^m \rangle = \sum_{\genotype_1,\ldots,\genotype_m} P_m(\genotype_1,\ldots,\genotype_m).
\label{Eq:momdef}
\end{equation}

For a genotype $\genotype_\alpha$ to be a local maximum, every $\mutvec_i$
has to satisfy the condition \cite{Hwang2017}
\begin{equation}
	\left |\phevec_\alpha + \left ( 1 - 2 \tau_i\right ) \mutvec_i \right | > \left | \phevec_\alpha \right |,
\end{equation}
where $\phevec_\alpha = \phevec(\genotype_\alpha)$ is the phenotype  associated with 
genotype $\genotype_\alpha$.
Defining the domain
\begin{equation}
\D[\phevec] = \left \{ \left . \y \in \mathbb{R}^n \right | 
|\y - \phevec | > |\phevec| \right \},
\label{Eq:area_def}
\end{equation}
we can succinctly write the condition for $m$ genotypes to be simultaneous local maxima as 
\begin{equation}
\mutvec_i \in \Area_i \equiv \bigcap_{\alpha=1}^m
	\D\left [ \left (2 \tau_{i,\alpha} -1 \right ) \phevec_\alpha\right ],
\end{equation}
where $\tau_{i,\alpha} \equiv \tau_i(\genotype_\alpha)$.
Using the definition of phenotype vectors \eref{Eq:gen_phe_map}, we get 
\begin{eqnarray}
\fl
\nonumber
P_m
= 
\int_{\mathbb{R}^n} \left ( \prod_{\alpha=1}^m \rmd  \phevec_\alpha \right )
	\left [ \prod_{j=1}^L \int_{\Area_j} \rmd  \mutvec_j p(\mutvec_j)  \right ]
	\prod_{\alpha=1}^m\delta\left ( \phevec_\alpha - \Q - \sum_{l=1}^L \mutvec_l \bn_{l,\alpha} \right )\\
\fl
=\int_{\mathbb{R}^n} \prod_{\alpha=1}^m \frac{\rmd  \phevec_\alpha \rmd  \vk_\alpha}{(2\pi)^{n}}
	\exp\left ( \rmi     \vk_\alpha \cdot (\phevec_\alpha-\Q)\right )
\left [ 
	\prod_{j=1}^L \int_{\Area_j} \rmd  \mutvec_j p(\mutvec_j)  
	\exp\left ( - \rmi   \mutvec_j\cdot \sum_{\beta=1}^m\vk_\beta\bn_{j,\beta} \right )
\right ],
\end{eqnarray}
where the Fourier representation of the delta function is used
and the arguments of $P_m$ are omitted for brevity.

Now we are ready to find a formal expression for the $m$th moment.
Using that 
\begin{eqnarray}
\nonumber
  \sum_{\genotype_1,\ldots,\genotype_m}
\prod_j \int_{\Area_j} \rmd  \mutvec_j p(\mutvec_j)  
	\exp\left ( - \rmi \mutvec_j\cdot \sum_{\beta=1}^m\vk_\beta\bn_{j,\beta} \right )\\
= \prod_{j=1}^L \left [ \sum_{\bn_{j,1}=0}^1\cdots
\sum_{\bn_{j,m}=0}^1   
\int_{\Area_j} \rmd  \mutvec_j p(\mutvec_j)  
	\exp\left ( - \rmi   \mutvec_j\cdot \sum_{\beta=1}^m\vk_\beta\bn_{j,\beta} \right )\right ],
\end{eqnarray}
we arrive at
\begin{equation}
  \label{Eq:Nm}
\langle \N^m \rangle
	=\int_{\mathbb{R}^n} \prod_{\alpha=1}^m \frac{\rmd  \phevec_\alpha \rmd  \vk_\alpha}{(2\pi)^{n}} \exp\left (\rmi \vk_\alpha \cdot (\phevec_\alpha-\Q)\right ) \left ( S_m\right )^L,
\end{equation}
where
\begin{equation}
S_m \equiv  \sum_{a_{1}=0}^1\cdots \sum_{a_{m}=0}^1 
	\int_{\Area(a)} \rmd  \mutvec \, p(\mutvec)  
\exp\left ( - \rmi   \mutvec\cdot \sum_{\beta=1}^m\vk_\beta a_\beta \right ),
\label{Eq:Sm}
\end{equation}
with  the domain of integration
\begin{equation}
	\Area(a) \equiv \bigcap_{\alpha=1}^m \D \left [  (2 a_\alpha-1) \phevec_\alpha \right ].
\end{equation}
In \ref{App:Moment_Cal}, we calculate $S_m$ and find that for large
$L$ 
\begin{equation}
\langle \N^m \rangle \approx \moment_m \left (\frac{2^L}{L^{1+n/2}} \right )^m
\exp\left ( - \frac{2m}{m+1} |\q|^2 L^{2\gamma-1} \right ),
\label{Eq:N_d}
\end{equation}
where $\moment_m$ is a constant independent of $L$ (see \eref{Eq:C_m} for the definition).
Within this derivation, the scaling of the wild-type phenotype was
chosen to be  of the form $\Q = \q L^\gamma$ with $0\le \gamma < 1$,
which implies that $q = \vert \q \vert$ can be treated perturbatively in the limit $L \to \infty$.
This approach is no longer valid if $\gamma =1$ and a separate analysis is required to determine $ \langle \N^m \rangle $. 
In~\cite{Hwang2017}, the nontrivial behaviour of $ \langle \N \rangle $ for $\gamma=1$ is discussed in detail.

In the following we consider the case $q=0$. The fact that $ \langle
\N^m \rangle $ is proportional to
$\left(\frac{2^L}{L^{1+n/2}}\right)^m$ suggests that the rescaled
random variable
\begin{equation}
  \label{Eq:scaledN}
\scaledN = \frac{L^{1+n/2}}{ 2^{L}} \N
\end{equation}
attains a nondegenerate limit distribution when $L \to \infty$. This
distribution will be explicitly computed for $n=1$ in the next
section. For general $n$, the scaling \eqref{Eq:scaledN} implies that
\begin{equation}
  \lim_{L \to \infty}  \frac{\ln \N}{L} = \ln 2
\end{equation}
on the level of single realizations. This shows that the exponential
growth rate defined in \eqref{Eq:Sigma_star} is $\Sigma^\ast = \ln 2$ in agreement
with the $\alpha \to 0$ limit of \eqref{Eq:SigmaFGM}, and moreover $\ln \N$
becomes a deterministic (self-averaging) quantity for $L \to \infty$.

The calculation presented in this section carries over in a very similar form to the local energy mininima of the AFHM
defined by the Hamiltonian \eqref{Eq:H_AFM} (see \ref{Sec:AFH}). The same scaling \eqref{Eq:scaledN} obtained for FGM applies, and the asymptotic expression
for the moments given in \eqref{Eq:Moments_AFHM} differs from \eqref{Eq:N_d} with $q=0$ by a factor $(m+1)^{n/2}$.

\section{\label{Sec:exact}Exact distribution in one-dimensional phenotype space}
In this section we limit ourselves to the one dimensional case with
$q=0$ and derive the probability density of the rescaled number
of fitness maxima in the large $L$ limit. 
Due to the simple geometry of one-dimensional Euclidean space, it is possible
to determine the exact form of the moments $\moment_m$ for $n=1$, from
which the full distribution can be extracted. 

\subsection{Probability density}
In \ref{Sec:Exact1d}, the $\moment_m$ for $n=1$ are obtained as 
\begin{equation}
\moment_m=
\frac{m! Q_m}{\sqrt{m+1}},\quad
Q_m \equiv 
\left [ \left ( \frac{1}{2} ; \frac{1}{2} \right )_m \right ]^{-1} ,
\label{Eq:1dtau}
\end{equation}
where we use the $q$-Pochhammer symbol defined by
\begin{equation}
  \label{Eq:Pochhammer}
(a;q)_m \equiv \prod_{k=0}^{m-1}(1-a q^k),
\end{equation}
with $(a;q)_0 \equiv 1$.
Some properties of the $q$-Pochhammer symbol are summarised in
\ref{Sec:QPoch}.
The $\moment_m$ are the moments of the rescaled random variable
\begin{equation}
  \label{Eq:X1D}
  X = \frac{L^{3/2}}{2^L} \N
\end{equation}
  defined in \eqref{Eq:scaledN} for general $n$, and we seek to derive
  the probability density $P(x)$ of $X$. 

We first consider the moment generating function $\G(k)$ of $X$ and its infinite
series representation 
\begin{equation}
\G(k) \equiv
	\int_{-\infty}^\infty P(x) \exp\left (\rmi kx\right ) \rmd x
= \sum_{m=0}^\infty \frac{\moment_m}{m!} (\rmi k)^m
= \sum_{m=0}^\infty \frac{Q_m(\rmi k)^m}{\sqrt{m+1}} ,
\label{Eq:GkSer}
\end{equation}
where we use \eref{Eq:1dtau}.
Because the radius of convergence of the infinite series is 1,
we need an analytic continuation to find
the probability density $P(x)$.

As we will see, $\G(k)$ can be written in terms of the Lerch
transcendent defined as~\cite{Erdelyi1955}
\begin{equation}
\wLi(z,s,v) \equiv \sum_{m=0}^\infty \frac{z^m}{(m+v)^s}.
\label{Eq:Lerch_def}
\end{equation}
Although $\wLi$ is defined for complex $s$ and $v$, we are only interested in the case where $v=1$ and $s$ is real
throughout this article. The third argument
of $\wLi$ will therefore be dropped in what follows.
The analytic continuation is obtained using 
the integral representation of $\wLi(z,s)$~\cite{Erdelyi1955}
\begin{equation}
\wLi (z,s) 
= \frac{1}{\Gamma(s)}\int_0^\infty \frac{t^{s-1} }{ \rme^t - z } \rmd t.
\end{equation}
If a branch cut is made from $z=1$ to $z=\infty$ along the real $z$ axis, 
$\wLi(z,s)$ is an analytic function in the cut plane for $s>0$.

Using \eref{Eq:Qcon} and \eref{Eq:Lerch_def}, we rewrite $\G(k)$ as
\begin{eqnarray}
\G(k) &= S \sum_{m=0}^\infty \frac{(\rmi k)^m}{\sqrt{m+1}}
\sum_{l=0}^\infty \frac{2^{-lm}}{(2;2)_l} \\
&= S\sum_{l=0}^\infty \frac{1}{(2;2)_l} \sum_{m=0}^\infty \frac{(\rmi k2^{-l})^m}{\sqrt{m+1}}
= S \sum_{l=0}^\infty \frac{\wLi(\rmi k 2^{-l},\frac{1}{2})}{(2;2)_l},\nonumber 
\label{Eq:G1}
\end{eqnarray}
where 
\begin{equation}
S \equiv \left [ \left (\frac{1}{2};\frac{1}{2}\right )_\infty \right ]^{-1} \approx 3.462\,7466.
	\label{Eq:S}
\end{equation}
Thus, we found a continuation of $\G(k)$ that is analytic in a Riemann sheet with a branch cut $\rmi k > 1$.

\begin{figure}[t]
\centering
\includegraphics[width=\linewidth]{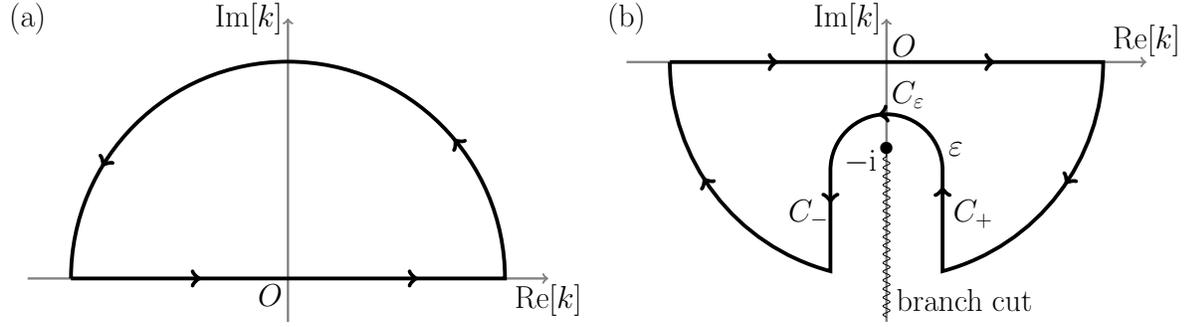}
\caption{\label{Fig:contour} Contour for the integral \eref{Eq:psi}. (a) Contour
	for negative $y$.
(b) Contour for positive $y$. 
	The branch point $-\rmi$ is indicated by a solid circle ($\bullet$) and
	the branch cut is indicated by a wiggly line. $C_+$ is the contour from $\varepsilon
	-\rmi \infty $ to $ \varepsilon -\rmi $ and $C_-$ is that from $-\varepsilon -\rmi$ to 
	$-\varepsilon - \rmi \infty$. The limit
        $\varepsilon \rightarrow 0$ is performed at the end of the calculation.
}
\end{figure}
Next, the probability density is obtained by the inverse Fourier transformation
\begin{eqnarray}
	\fl
	P(x) &= \frac{1}{2\pi} \int_{-\infty}^\infty \exp\left (-\rmi kx\right ) \G(k) \rmd k  \nonumber \\
	\fl
	&= \frac{S}{2\pi} \sum_{l=0}^\infty \frac{1}{(2;2)_l} \int_{-\infty}^\infty \exp\left (-\rmi kx\right ) 
\wLi\left ( \rmi  k2^{-l}  ,\frac{1}{2}\right ) \rmd k 
=S \sum_{l=0}^\infty \frac{2^l }{(2;2)_l} \psi\left (2^l x\right ),
	\label{Eq:PN}
\end{eqnarray}
where
\begin{equation}
	\psi(y) \equiv \frac{1}{2\pi} \int_{-\infty}^\infty \exp\left (-\rmi ky\right ) \wLi\left (\rmi k,\frac{1}{2} \right ) \rmd k.
\label{Eq:psi}
\end{equation}
For $y<0$, we consider the contour in
the complex $k$ plane shown in \fref{Fig:contour}~(a). 
Since $\wLi(\rmi k)$ has a branch point at $\rmi k=1$ and a branch cut $\rmi k>1$ [see
\fref{Fig:contour}~(b)], the contour integral gives $\psi(y) = 0$. 
Thus, $P(x) = 0$ for $x<0$ as it should be.

For positive $y$, we consider the contour in
\fref{Fig:contour}~(b). Since $\Phi(z,\frac{1}{2}) \sim 1/\sqrt{1-z}$ for $|1-z|\ll 1$~\cite{Erdelyi1955},
the integral over $C_\varepsilon$ approaches zero as $\varepsilon \rightarrow 0$.
Hence,  the nonzero contribution to the integral comes from the contours $C_+$ and $C_-$:
\begin{eqnarray}
\psi(y) = -\frac{1}{2\pi}\lim_{\varepsilon \rightarrow 0} \left [ \int_{C_+} + \int_{C_-} 
	\right ] \exp\left (-\rmi zy\right ) \wLi\left (\rmi    z,\frac{1}{2} \right ) \rmd z
	\nonumber \\
	=\frac{1}{2\pi i}\lim_{\varepsilon \rightarrow 0}\int_1^\infty \rme^{-yw}
	\left [ 
	\rme^{-\rmi \varepsilon y} \wLi\left( w + \rmi \varepsilon ,\frac{1}{2}\right )
	-\rme^{\rmi \varepsilon y} \wLi\left( w - \rmi \varepsilon ,\frac{1}{2}\right )
	\right ]\rmd w \nonumber \\
	=\frac{1}{\pi} \lim_{\varepsilon\rightarrow 0}
	\int_1^\infty \rme^{-yw} \Im \wLi\left( w + \rmi \varepsilon ,\frac{1}{2}\right )\rmd w ,
\label{Eq:psi_C}
\end{eqnarray}
where $\Im z$ stands for the imaginary part of $z$ and
we have used $\wLi(z,s)^* = \wLi(z^*,s)$ (the asterisk represents complex conjugation).
Using
\begin{equation}
\lim_{\varepsilon\rightarrow 0} \frac{1}{x- \rmi   \varepsilon} 
= \frac{1}{x} + \rmi   \pi\delta (x),
\end{equation}
we obtain
\begin{equation}
\Im\wLi\left (w+ \rmi \varepsilon,\frac{1}{2} \right ) = \Im\int_0^\infty \frac{\rmd t}{\sqrt{\pi t} (\rme^t - w -\rmi   \varepsilon)}
	=  \frac{\sqrt{\pi}}{w\sqrt{\ln w}},
\end{equation}
which gives
\begin{equation}
	\psi(y) = \frac{1}{\sqrt{\pi}} \int_1^\infty \frac{\exp\left (-yw\right )}{w \sqrt{\ln w}}\rmd w
=  \int_0^\infty \frac{\exp(-y\rme^t)}{\sqrt{\pi t}} \rmd t.
\label{Eq:psix}
\end{equation}
In \ref{App:Another}, we derive the same distribution using
a slightly different method.

\begin{figure}[t]
\centering
\includegraphics[width=0.7\textwidth]{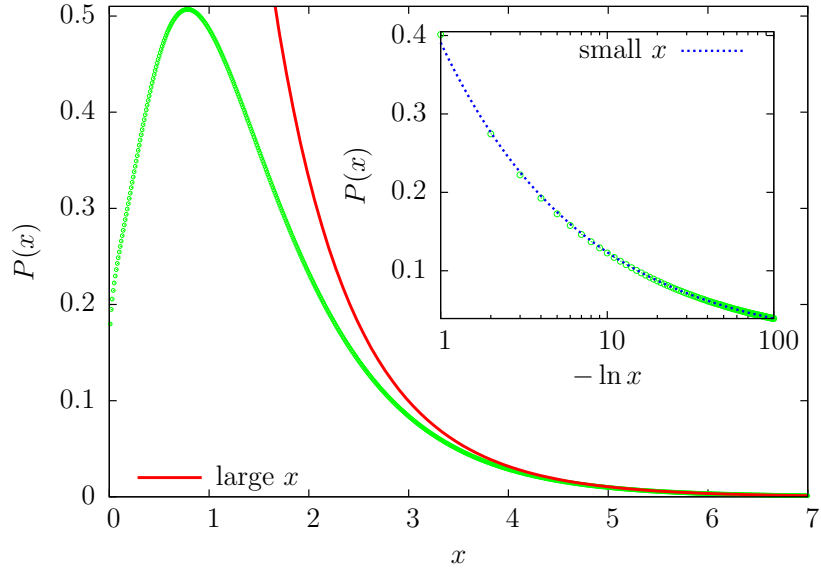}
	\caption{\label{Fig:prob} Plot of the probability density
          $P(x)$ of the scaled number of fitness maxima for FGM with $n=1$ (green open circles). 
	For comparison, the leading asymptotic behaviour \eref{Eq:Pas} for large $x$ is drawn
as a red curve. Inset: semi-logarithmic
	plot of $P(x)$ vs. $-\ln x$ illustrating the behaviour \eref{Eq:Pas} for small
$x$. A curve showing the leading asymptotics is drawn for comparison
(dotted blue line).
}
\end{figure}

In \fref{Fig:prob}, we depict $P(x)$ obtained by numerical evaluation
of \eref{Eq:psix} and \eref{Eq:PN}. 
One may observe that $P(x)$ seems to approach a nonzero value as $x\rightarrow 0$.
A careful analysis presented in \ref{App:Asym} shows, however, 
that $P(x) \rightarrow 0$ as $x\rightarrow 0$ with an infinite slope.
For large $x$, $P(x)$ is dominated by the leading order $l=0$ term in \eqref{Eq:PN} and the asymptotics reflect that of $\psi(x)$. 
Taken together, the behaviour of $P(x)$ for large and small $x$ is found to be
\begin{equation}
P(x) \sim 
	\begin{cases}
		\displaystyle \frac{\ln 2}{\sqrt{-\pi \ln x}}, & \text{ for } x \ll 1,\\
		\displaystyle S \frac{\rme^{-x}}{\sqrt{x}}, &\text{ for } x \gg 1.
	\end{cases}
\label{Eq:Pas}
\end{equation}
The asymptotic behavior is compared to the exact probability density in the inset of \fref{Fig:prob}.

\subsection{Finite $L$ correction}
\label{Sec:FiniteSize}

To facilitate the comparison to numerical simulations, we consider 
the finite-$L$ corrections to the distribution $P(x)$.
In \ref{Sec:Sub} we obtain the $\Or(1/L)$ correction to the moments of
the rescaled variable $X$ as
\begin{equation}
\frac{\langle \scaledN^m \rangle 
 - \moment_m}{m!} =  - \frac{3}{4L} 
\frac{ m^2(m+2)Q_m}{ (m+1)^{3/2}} + \Or(L^{-2}).
\end{equation}
Writing the moment generating function of $\scaledN$ for finite $L$ as $\G(k) + \Delta \G(k)$,
we get
\begin{eqnarray}
-\frac{4L}{3} \Delta \G &\approx  \sum_{m=0}^\infty \frac{m^2(m+2)}{(m+1)^{3/2}} Q_m
(\rmi k)^m
\nonumber
\\
\nonumber
&= \sum_{l=1}^\infty Q_{l-1}\left ( l^{3/2} - l^{1/2} 
- l^{-1/2}
+ l^{-3/2} \right ) (\rmi k)^{l-1} \\
&\equiv \G_{3/2} - \G_{1/2} - \G_{-1/2} + \G_{-3/2},
\end{eqnarray}
where
\begin{equation}
\G_{s}(k) = \sum_{l=1}^\infty Q_{l-1}  (\rmi k)^{l-1}l^{-s}
= S \sum_{m=0}^\infty \frac{\wLi(\rmi k2^{-m},s)}{(2;2)_m}.
\end{equation}
Note that $\G_{1/2}(k) = \G(k)$.

\begin{figure}[t]
\centering
\includegraphics[width=\textwidth]{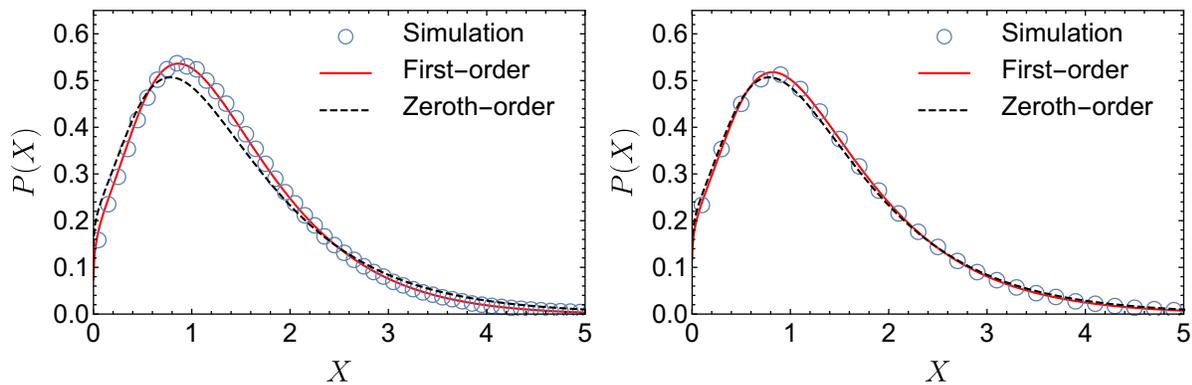}
\caption{\label{Fig:oneD} Probability density of 
the scaled number of fitness maxima $\scaledN$ for $L=20$ (left panel)
and $L=50$ (right panel). Simulation results are compared to the
	asymptotic density $P(x)$~\eref{Eq:PN} and the first-order $\Or(1/L)$ correction~\eref{Eq:DPx}.}
\end{figure}

If we denote the Fourier transform of $\wLi(\rmi k,s)$ by
\begin{equation}
	\psi_s(x) \equiv \frac{1}{2 \pi} \int_{-\infty}^\infty \exp\left (-\rmi kx\right )
\wLi(\rmi k,s) \rmd k,
\end{equation}
we obtain a recursion relation 
\begin{eqnarray}
\psi_{l-1}(x) &= \frac{1}{2\pi} \int_{-\infty}^\infty
	\exp\left (-\rmi kx\right ) \wLi(\rmi k,l-1) \rmd k\nonumber \\
&= \frac{1}{2\pi} \int_{-\infty}^\infty
	\exp\left (-\rmi kx\right ) \frac{\rmd}{\rmd k} \left [ k \wLi(\rmi k,l)\right ] \rmd k
= \left ( - x \frac{\rmd }{\rmd x} \right ) \psi_l(x),
\end{eqnarray} 
where we have used
\begin{equation}
\wLi(z,l-1) = \frac{\rmd }{\rmd z} \left [ z \wLi(z,l) \right ].
\end{equation}
Since $\psi_{1/2}(x)=\psi(x)$ in \eref{Eq:psix}, 
we have
\begin{eqnarray}
\nonumber 
\psi_{-1/2}(x) 
&= \frac{x}{\sqrt{\pi}}  \int_0^\infty \frac{\exp(t-x\rme^t)}{\sqrt{t}} \rmd    t,\\
\psi_{-3/2}(x) &= -\psi_{-1/2}(x) 
+ \frac{x^2}{\sqrt{\pi}} \int_0^\infty \frac{\exp(2 t-x\rme^t)}{\sqrt{t}} \rmd    t.
\end{eqnarray}
To find $\psi_{3/2}(x)$, we use the integral representation for $\wLi(z,\frac{3}{2})$,
\begin{equation}
\wLi\left (z,\frac{3}{2} \right )= \Gamma\left (\frac{3}{2}\right )^{-1} \int_0^\infty \frac{\sqrt{t}}{\rme^t - z} \rmd    t
\end{equation}
and perform the contour integral along the contour in \fref{Fig:contour},
which gives
\begin{equation}
	\psi_{3/2}(x) = \frac{2}{\sqrt{\pi}} \int_1^\infty \frac{\exp\left (-wx\right )\sqrt{\ln w}}{w}\rmd      w
= \frac{2}{\sqrt{\pi}} \int_0^\infty \sqrt{t} \exp(-x\rme^t) \rmd    t.
\end{equation}
Hence the correction to the probability density $P(x)$ is given by 
\begin{equation}
 \Delta P(x) = -\frac{3S}{4L} \sum_{l=0}^\infty \frac{\Psi(x2^{l})}{(2;2)_l},
	\label{Eq:DPx}
\end{equation}
where
\begin{equation}
\Psi(x) = \frac{1}{\sqrt{\pi}}
\int_0^\infty \rmd    t \frac{\exp(-x \rme^t)}{\sqrt{t}} 
\left ( 2 t -1 - 2 x \rme^t + x^2 \rme^{2t} \right ).
\end{equation} 
In \fref{Fig:oneD}, we compare our prediction with simulations,
which shows an excellent agreement already for $L=20$. The simulation
method is explained in \ref{Sec:Simul}. 

\subsection{One-dimensional AFHM and the number partioning problem}

\begin{figure}
	\includegraphics[width=\linewidth]{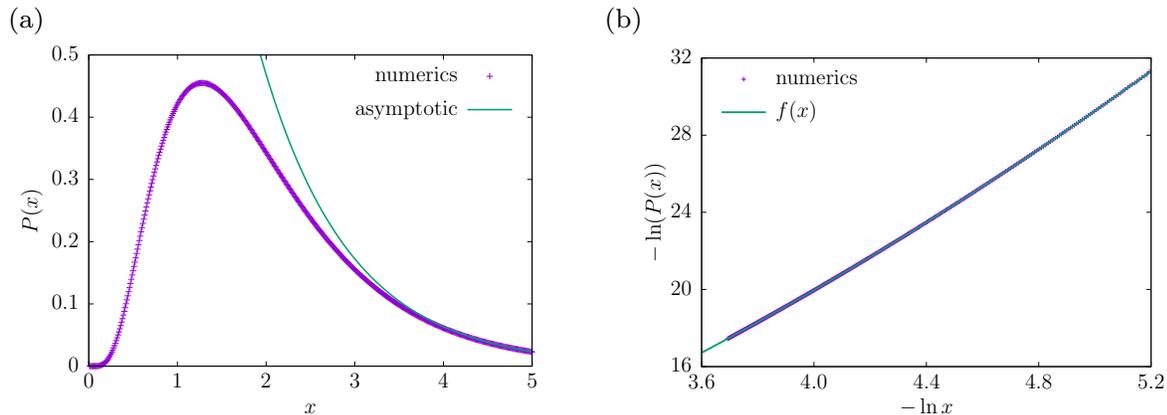}
	\caption{\label{Fig:AFN}(a) Probability density $P(x)$ of the
          scaled number of fitness maxima for the AFHM with $n=1$ (symbols) 
	together with the exponential behaviour $S e^{-x}$ for large $x$ (line), where
	$S$ is defined in \eref{Eq:S}.
	The function $P(x)$ is obtained by numerical summation of
        \eqref{Eq:P_AFHM}. 
	(b) Plot of $-\ln(P(x))$ vs. $-\ln x$ (symbols). 
	The fitting function $f(x)=a x^2 + b x + c$ with $a=0.851$,
        $b=1.64$, $c=-0.215$ (line) is
	almost indistinguishable from the numerical data (see \ref{Sec:AFH}
        for details).}
\end{figure}

In \ref{Sec:AFH} the calculation of the probability density is repeated for the one-dimensional AFHM, and 
the limiting distribution is found to be
\begin{equation}
  \label{Eq:P_AFHM}
	P_\mathrm{AFHM}(x) =S \sum_{l=0}^\infty \frac{2^l }{(2;2)_l} \exp\left (-2^l x\right ).
\end{equation}
Again the behaviour for large $x$ is determined by the $l=0$ term and is simply exponential in this case. However, the behaviour for small $x$ differs markedly from
that of FGM. In fact the expression \eqref{Eq:P_AFHM} can be shown to have vanishing derivatives of all orders at $x=0$, which implies an essential singularity
at the origin (\fref{Fig:AFN}). Thus, whereas small values of $X$ are relatively likely for FGM, they are very rare in the AFHM.

The one-dimensional AFHM is closely related to the number partioning
problem (NPP) \cite{Ferreira1998,Mertens2000}. In this problem one asks for the
optimal subdivision of $L$ positive random numbers $\xi_i$,
$i=1,\dots,L$ into two subsets ${\cal{S}}_1, {\cal{S}}_2$ such that
the difference $\Delta$ between the sums of the
$\xi_i$ over the subsets is as small as possible. Setting $s_i = 1$ if
$i \in {\cal{S}}_1$ and $s_i = -1$ if $i \in {\cal{S}}_2$ the
difference can be written as
\begin{equation}
  \Delta = \sum_{i \in {\cal{S}}_1} \xi_i - \sum_{j \in {\cal{S}}_2}
  \xi_j = \sum_{i=1}^L \xi_i s_i,
\end{equation}
and $\vert \Delta \vert^2$ is seen to be proportional to the
one-dimensional AFHM Hamiltonian. In \cite{Ferreira1998} the expected
number of local minima of $\vert \Delta \vert^2$ was computed for the case when
the $\xi_i$ are uniform random variable on the interval $[0,1]$. The
result
\begin{equation}
 \langle \N \rangle_\mathrm{NPP} \sim \sqrt{\frac{24}{\pi}}
 \frac{2^L}{L^{3/2}}, \;\;\; L \to \infty,
\end{equation}
displays the same scaling with $L$ that we have obtained for FGM and
AFHM. The prefactor can be obtained from our result \eqref{Eq:Mom_AFHM} for
the AFHM using the rescaling \eqref{Eq:mom_gen} with $\omega^2 =
\frac{1}{12}$ and $p(0) = 1$ for the uniform distribution. 

\section{Correlation between local maxima}
\label{Sec:correlations}
In this section we consider the conditional probability $P(\genotype_2|\genotype_1)$ that
a genotype $\genotype_2$ is a local maximum, given that
$\genotype_1$ is also a local maximum, for FGM with a one-dimensional
phenotype space. This is to
be compared to the unconditional probability $P_1(\sigma_2)$ that
$\genotype_2$ is a local maximum.
Using the notation in \sref{Sec:moments}, we define
\begin{equation}
        C(\genotype_1,\genotype_2) 
	= \frac{P(\genotype_2|\genotype_1)}{P_1(\genotype_2)}
	= \frac{P_2(\genotype_1,\genotype_2)}{P_1(\genotype_1)P_1(\genotype_2)}.
\end{equation}
Due to permutation symmetry, $P_2$ depends only on the 
following four parameters:
\begin{eqnarray}
	\nonumber
	u_0\equiv \sum_{i=1}^L  \left [ 1 - \bn_i(\sigma_1)\right ]
	\left [ 1 - \bn_i(\sigma_2) \right ],\quad
	&u_1\equiv \sum_{i=1}^L  \left [ 1 - \bn_i(\sigma_1)\right ]\bn_i(\sigma_2),\\
	u_2\equiv \sum_{i=1}^L \bn_i(\sigma_1) \left [ 1 - \bn_i(\sigma_2)\right ],\quad
	&u_3\equiv \sum_{i=1}^L \bn_i(\sigma_1) \bn_i(\sigma_2).
	\label{Eq:s0s3}
\end{eqnarray}
Obviously, $u_0 + u_1 + u_2 + u_3 = L$.
These parameters can be interpreted as follows:
$u_0$ is the number of shared non-mutated sites (i.e., the number of $00$ pairs in
a sequence alignment),
$u_3$ is the number of shared mutated sites ($11$ pairs),
$u_1$ is the number of sites that do not have mutations in $\genotype_1$ but
have mutations in $\genotype_2$ ($01$ pairs), and
$u_2$ is the number of sites that do not have mutations in $\genotype_2$ but
have mutations in $\genotype_1$ ($10$ pairs); see \eref{Eq:align} for a pictorial representation.
As shown in \ref{App:TwoPoint}, for large $u_i$ the probabilities
$P_1$ and $P_2$  can be approximated as
\begin{eqnarray}
	P_1(\genotype_1) = \frac{1}{L\sqrt{d_1}},\quad
	P_1(\genotype_2) = \frac{1}{L\sqrt{d_2}},
	\label{Eq:P1}
	\\
	\label{Eq:P2}
	P_2(\genotype_1,\genotype_2) = \frac{3}{\left(L+d_{12} \right)
  \left(2L - d_{12}\right) \left ( d_1 d_2 - u_3^2 \right )^{1/2}},
\end{eqnarray}
which yields
\begin{eqnarray}
	C(\genotype_1,\genotype_2) = 3 
	\left ( 1 +\frac{d_{12}}{L} \right )^{-1}
	\left ( 2 -\frac{d_{12}}{L} \right )^{-1}
	\left ( 1 -\frac{u_3^2}{d_1 d_2} \right )^{-1/2}
\label{Eq:TwoPointFin}
\end{eqnarray}
with $d_1=u_2+u_3$, $d_2=u_1+u_3$, and $d_{12} = u_1+u_2$.
Here $d_i$ is the Hamming distance from the wild type to $\genotype_i$
and 
$d_{12}$ is the Hamming distance between $\genotype_1$ and $\genotype_2$.

To discuss the significance of \eref{Eq:P1}, \eref{Eq:P2}, and \eref{Eq:TwoPointFin}, we first consider
two genotypes with $d_1/L \approx d_2/L \approx d_{12}/L \approx \frac{1}{2}$ for large $L$,
or   
$u_i/L \approx \frac{1}{4}$ for $i = 0, 1, 2, 3$. For this set of values, we get
\begin{equation}
	P_1^* = \sqrt{2}L^{-3/2} = \mu_1 L^{-3/2} = \frac{\langle \N
          \rangle}{2^L}, \quad
	P_2^* = \frac{16}{\sqrt{27} L^3} = \mu_2  L^{-3} =
        \frac{\langle \N^2 \rangle}{2^{2L}}.
\end{equation}
This shows that a local maximum is typically located around $d= L/2$
and similarly a typical pair of local maxima is separated by Hamming
distance $d_{12} = L/2$, as would be expected for entropic reasons.
For two randomly chosen genotypes we therefore have  
\begin{equation}
	C^\ast \equiv \frac{P_2^*}{\left ( P_1^* \right )^2}
	=\frac{\mu_2}{\mu_1^2}
	= \frac{8}{\sqrt{27}} \simeq 1.54 > 1
      \end{equation}
      simply because the distribution of the scaled number of maxima has a
      nonzero width. 

Next we observe that when $u_3 = 0$ (no shared mutations), $C$ takes
on its minimal value $\frac{4}{3}$ when $d_{12} = \frac{1}{2} L$. As $C$ is an increasing function of $u_3$
for fixed $d_{12}$, this constitutes a global lower bound on $C$,
\begin{equation}
	C(\genotype_1,\genotype_2) \ge C_\text{min} =  \frac{4}{3} < C^\ast.
\end{equation}
Two randomly chosen genotypes conditioned to have no shared mutations are thus less likely to be maxima
than expected for unconstrained sequences. 

\begin{figure}[t]
\centering
\includegraphics[width=\linewidth]{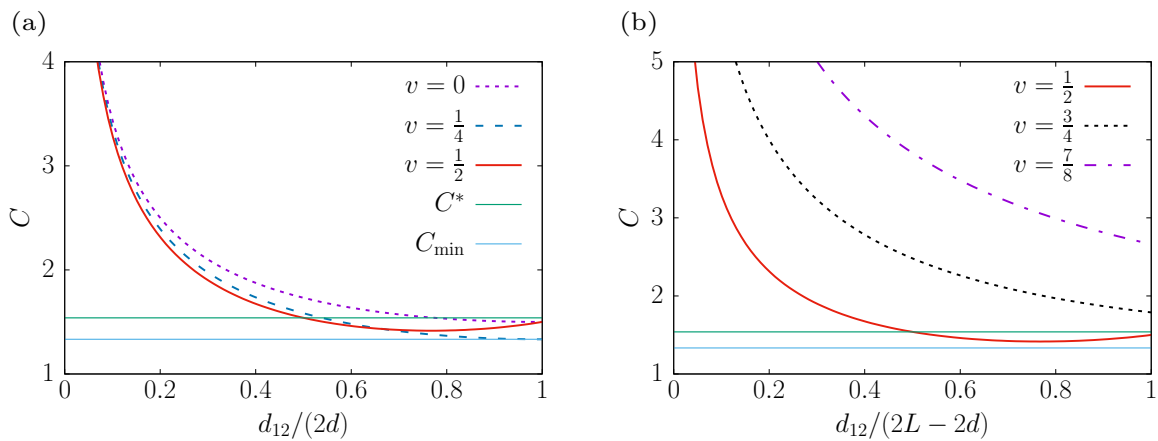}
\caption{\label{Fig:coor} Correlation between two local maxima located
at the same distance $d_1 = d_2 = d$ from the wild type. (a)
$C(\genotype_1,\genotype_2)$ as a function of $d_{12}/(2d)$ for different values
	of $v = d/L \leq 1/2$; see \eref{Eq:Cs}. Since we are considering the infinite $L$
  limit, $v=0$ does not necessarily mean
	$d=0$, but rather that $d$ is small compared to $L$ (for example, $d=\sqrt{L}$). (b) $C(\genotype_1,\genotype_2)$ as a function of $d_{12}/(2L-2d)$ for different values
	of $v = d/L \geq 1/2$; see~\eref{Eq:Cl}. In both panels the random expectation
  $C^\ast$ and the lower bound $C_\text{min}$ are depicted as
  horizontal lines.}
\end{figure}

It is also instructive to analyse the symmetric case $d_1=d_2 = d$,
where both genotypes are at the same distance from the wild type.  
Since $d_{12} \leq 2\min(d,L-d)$, we choose $w \equiv d_{12}/[2 \min(d,L-d)]$ as our free parameter. 
In terms of $w$, $C$ can be written as
\begin{eqnarray}
	\fl
	\label{Eq:Cs}
	C = \frac{3}{2 \sqrt{w} (1+2vw)(1-vw) \left( 2 - w \right)^{1/2}},&\text{ for } v \le \frac{1}{2},\\
	\fl
	\label{Eq:Cl}
	C = \frac{3v}{2\sqrt{w(1-v)}[1+2(1-v)w][1-(1-v)w][(2+w)v-w]^{1/2}},&\text{ for } v \ge \frac{1}{2},
\end{eqnarray}
where $v \equiv d/L$. The divergence for $w \ll 1$ shows that nearby
maxima are clustered in sequence space, an effect that has been found
also in other fitness landscape models \cite{Nowak2015}. Nevertheless
there are regions where maxima effectively repel, in the sense
that $C$ is smaller than the random expectation $C^\ast$, and moreover
the correlations do not always vary monotonically with
$d_{12}$ (\fref{Fig:coor}).

\section{Summary and discussion}
\label{Sec:summary}
In this paper, we studied the distribution of the number $\N$ of local maxima 
in the genotypic fitness landscapes generated according to Fisher's geometric model (FGM) with 
phenotypic dimension $n$.
We first examined the connection between FGM and the anti-ferromagnetic
Hopfield model (AFHM) with $n$ real-valued patterns, where local
fitness maxima correspond to zero-temperature metastable states that are stable under single spin flips. When the
phenotypic dimension $n$ and the genotype sequence length $L$
(corresponding to the number of spins in the AFMH) are jointly taken
to infinity at fixed but small ratio $\alpha = n/L$, we find
that the exponential growth rate $\Sigma^*$ of the mean number of
maxima is identical for the two models up to $\Or(\alpha \ln |\ln \alpha|)$. 

More detailed results are obtained when the limit $L \to \infty$ is
performed at finite $n$. In this case, we show that 
$X = \N L^{1+n/2}/2^L$ is an appropriate rescaled random variable with a well-defined probability
density both for FGM and the AFHM. In particular,
we derive the exact probability densities $P(X)$ for both models in
the case $n=1$. Despite the identical scaling, the two densities display remarkably different
behaviours for small $X$. 
Furthermore, we compute the leading finite size correction to the
distribution and show that the obtained analytic expression
agrees well with simulation results.
Finally, we provide a detailed analysis of the pairwise correlations between
the positions of local fitness maxima in the one-dimensional FGM,
finding a pronounced clustering of maxima at small Hamming
distance. To the best of our knowledge these are the first
analytic results for the correlation between maxima in a fitness
landscape model with nontrivial structure.

The full distribution of fitness maxima has been found only in
a few fitness landscape models so far, but already this small number of examples
suggests a diverse range of possible scenarios. The simplest genotypic fitness landscape is
the House-of-Cards (HoC) model, where fitness values are drawn from a
continuous probability distribution and assigned independently to
genotypes \cite{Kingman1978,Kauffman1987}; the corresponding spin system is known as the Random
Energy model \cite{Derrida1981}. In the HoC model the
distribution converges to a Gaussian for large $L$, with a variance that is
proportional to the mean \cite{Macken1989,Baldi1989}. This implies
that the number of maxima $\N$ itself becomes a deterministic
(self-averaging) quantity.

Another solvable case is the NK block
model, where the $L$ sites of the sequence are subdivided into
disjoint subsets of size $k$. The fitness landscape of each subset
is an uncorrelated HoC landscape, and the fitness of the genotype is the sum of the contributions
of the subsets \cite{Hwang2018,Perelson1995}. The total number of fitness maxima is then the
product of the numbers of maxima of the sublandscapes, and therefore
the distribution of $\N$ becomes log-normal in the limit $L \to
\infty$ at fixed $k$ \cite{Schmiegelt2014}. As a consequence $\ln \N$
is self-averaging, but a scaling form for $\N$ similar to that found
here for FGM does not exist, because the moment $\langle \N^m
\rangle$ does not scale as the $m$th power of $\langle \N \rangle$. It
would be of interest to investigate the limiting distribution of the
number of maxima that arises in this model (as well as in other
versions of the NK model \cite{Hwang2018}) when the joint limit $k,L
\to \infty$ is performed at fixed ratio $k/L$.

Yet stronger fluctuations in $\N$ are found in FGM when the distance of the wild-type phenotype to the fitness optimum is nonzero
and scales as $\vert \Q \vert = q L$. In \cite{Hwang2017} the
exponential growth rate $\Sigma^\ast$ of the mean number of maxima was
computed as a function of $q$, and was found to vanish at $q_c \approx
0.924809$. On the other hand, the typical value of $\ln \N$ can be
obtained from a thermodynamic calculation of the entropy of the
model~\cite{HwangUn},
which shows that the extensive part of $\langle \ln \N \rangle$ vanishes already at $q =
\frac{1}{\sqrt{2\pi}} \approx 0.399$. Thus
for $0 < q < q_c$, $\ln \langle \N \rangle \gg \langle \ln \N \rangle$
and the self-averaging property breaks down also on the level of $\ln
\N$. Preliminary work on the thermodynamics of the model for general $\alpha$ suggests that this glassy behavior is
typical throughout the $\alpha-q$-phase diagram, such that $\ln \N$ is self-averaging only at the point $\alpha = q = 0$. 

From a biological perspective it is of interest to go beyond the assumption of binary genotype sequences and consider
models where the number of possible states per site (the number of \textit{alleles}) is $A > 2$ \cite{Zagorski2016,Schmiegelt2019}.
This modification has opposing effects on the number of fitness maxima. On the one hand, the total number of genotypes
increases trivially to $A^L$, but at the same time the number of conditions that have to be satisfied for a genotype to be a
fitness peak also increases. For the HoC model \cite{Kauffman1987} and the NK block model \cite{Schmiegelt2014} these effects
are easily accounted for. However, for FGM the analysis of the multiallelic generalization proposed in \cite{Hwang2017} is
highly nontrivial and will be presented elsewhere \cite{HwangUn2}. 

To conclude, FGM is a paradigm for understanding how complex genotypic fitness landscapes
arise from combining a simple (linear) genotype-phenotype map with an equally simple (nonlinear but single-peaked)
phenotype-fitness map \cite{Domingo2019,Manrubia2020}. This paradigm is becoming increasingly relevant for the analysis of
large-scale empirical data sets encompassing hundreds of thousands of genetic sequences
\cite{Pokusaeva2019}. Sample-to-sample fluctuations in summary statistics such as the number of fitness peaks constitute
a significant obstacle to inference methods aimed at extracting low-dimensional phenotypes from genotype-fitness data
\cite{Blanquart2016}. We hope that the present case study can help to address this problem and contribute to the further
development of fitness landscape methods in evolutionary genetics. 

\ack
SCP acknowledges support by the Basic Science Research Program through
the National Research Foundation of Korea (NRF) funded by the Ministry
of Science and ICT (Grant No. 2017R1D1A1B03034878); and by the Catholic University of Korea,
research fund 2019. SH and JK
acknowledge support by DFG within CRC 680 and CRC 1310.
The authors furthermore thank the Regional Computing Center of the
University of Cologne (RRZK) for providing computing time on the DFG-funded High
Performance Computing (HPC) system CHEOPS.

\appendix
\section{\label{App:Q0} Small $\alpha$ limit for $\Q=0$}
As discussed in \sref{Sec:HM}, FGM with $\Q=0$ can be mapped into a
certain variant of the AFHM.  
Here, we perform a direct comparison between the two models in terms of
the exponential growth rate of the mean number of maxima $\Sigma^*$ defined in \eref{Eq:Sigma_star}.
In our recent study of FGM~\cite{Hwang2017}, we have shown that $\Sigma^*$ is 
obtained by maximizing the function $\Sigma(a,b,c)$ with respect to
$a$, $b$, and $c$\footnote{The original equation (46) in
  \cite{Hwang2017} used an alternative variational parameter $g$ which
  is defined as $ 16 q^2 c = \alpha^2 - g^2$. However, in our setting
  $q=0$, and it is natural to use $c$ since $g$ is simply $\alpha$.}, where
\begin{eqnarray}
\Sigma(a,b,c) =
-\frac{\alpha}{2}\ln \left(\frac{\alpha^2 }{ a c+b^2}\right)+\alpha + b -2c 
	+\ln M,
\label{Eq:Action}\\
	M \equiv \frac{1}{2}\left [ \mathrm{erf}\left(\frac{\alpha +2 b}{\sqrt{2 a}}\right)+1 \right ]+\frac{\rme^{2 c}}{2} \left[\mathrm{erf}\left(\frac{\alpha }{\sqrt{2 a} }\right)+1\right ],
	\label{Eq:M}
\end{eqnarray}
and $\mathrm{erf}(x)$ is the error function.
By taking derivatives with respect to each variable, we get
\begin{eqnarray}
	\frac{\alpha  c}{2 \left(a c+b^2\right)}-\frac{(\alpha + 2 b) X + \alpha Y  }{2 M a \sqrt{2\pi a}}=0,\label{Eq:a}\\
		1+\frac{\alpha  b}{a c+b^2}+\left ( \frac{2}{a\pi} \right )^{1/2}\frac{X}{M}=0,\label{Eq:b}\\
			\frac{a \alpha }{2 \left(a c+b^2\right)}-\frac{1}{M} \left[\text{erf}\left(\frac{\alpha +2 b}{\sqrt{2a}}\right)+1\right]=0,	
			\label{Eq:c}
\end{eqnarray}
where
\begin{equation}
	X\equiv \exp \left (-\frac{(\alpha +2 b)^2}{2 a}\right ),\quad
	Y\equiv \exp \left (2c -\frac{\alpha^2}{2 a}\right ).
\end{equation}
The solution $(a,b,c) = (a^*,b^*,c^*)$ of 
\eref{Eq:a}, \eref{Eq:b}, and \eref{Eq:c}
determines $\Sigma^* = \Sigma(a^*,b^*,c^*)$.

To find an approximate solution, we first observe that
$\Sigma^* = \ln 2$ for $\alpha = 0$ according to \eref{Eq:N_d}.
Thus for $\alpha \rightarrow 0$,
$M$ should approach $2$ and
the arguments of both error functions in \eref{Eq:M} should diverge, 
which suggests (we drop the asterisks for brevity)
\begin{equation}
	a = \frac{1}{2} \alpha^2 A(\alpha),
	\label{Eq:a_app}
\end{equation}
with $A(\alpha) \rightarrow 0$ as $\alpha \rightarrow 0$. 

From \eref{Eq:c} together with the above observation, we get
\begin{equation}
	\frac{a \alpha}{a c + b^2} = 2 + \sor(1)
	\rightarrow c + \frac{b^2}{a} = \frac{1}{2} \alpha + \sor(\alpha),
	\label{Eq:c_con}
\end{equation}
from  which we conclude that $|b| \ll \alpha$ (accordingly, $\alpha + 2 b \approx \alpha$) and $c\ll 1$. 
Note that because of \eref{Eq:a} $c$ is positive.
Therefore, we have
\begin{equation}
	X \approx Y \approx \exp \left ( -\frac{1}{A} \right ).
	\label{Eq:XY}
\end{equation}
Using \eref{Eq:c_con} and \eref{Eq:XY}, we can approximate
\eref{Eq:a} and \eref{Eq:b} as
\begin{equation}
	c \approx \frac{\exp(-1/A)}{(4\pi A)^{1/2}},
	\quad
	1+\frac{2 b}{a} + \frac{2c}{\alpha} \approx 0.
	\label{Eq:final}
\end{equation}
Since $c$ is at most $\Or(\alpha)$, $b$ must be $\Or(a)$,
which, along with \eref{Eq:c_con} and \eref{Eq:a_app}, gives
\begin{equation}
	c \approx \frac{1}{2} \alpha.
\end{equation}
Thus, \eref{Eq:final} yields $b \approx -a$ and 
\begin{equation}
	\frac{\exp(-1/A)}{(\pi A)^{1/2}} \approx \alpha.
	\label{Eq:Aa}
\end{equation}
Using successive approximations to solve \eqref{Eq:Aa}, we get
\begin{equation}
	\frac{1}{A}= \ln \frac{1}{\alpha\sqrt{\pi} } + \frac{1}{2} \ln \frac{1}{A}  \approx \ln \frac{1}{ \alpha\sqrt{\pi}} + \frac{1}{2} \ln \left ( \ln \frac{1}{\alpha\sqrt{\pi}}  \right ).
	\label{Eq:Aapprox}
\end{equation}

To find the asymptotic behaviour of $\Sigma^*$ for small $\alpha$,
we exploit the asymptotics of the error function 
$\mathrm{erf}(x) \sim 1 - \exp(-x^2)/(\sqrt{\pi} x)$ to write
\begin{equation}
	\mathrm{erf}\left ( \frac{\alpha + 2b}{\sqrt{2a}} \right )
	\simeq
	\mathrm{erf}\left ( \frac{\alpha}{\sqrt{2a}} \right )
	= \mathrm{erf} \left ( \frac{1}{\sqrt{A}} \right )
	\sim 1 - A \frac{\exp\left (-1/{A} \right ) }{\left (\pi A\right )^{1/2}}
	\approx 1 - \alpha A,
\end{equation}
where we use \eref{Eq:Aa}. We can now approximate $M$ as
\begin{equation}
	M \approx \left ( 2 - \alpha A \right ) \frac{1+\rme^{2c}}{2}
	\approx \left ( 2 - \alpha A \right ) \left ( 1 + \frac{\alpha}{2} \right ) 
	\approx 2 + \alpha - \alpha A.
	\label{Eq:Mapprox}
\end{equation}
Using \eref{Eq:c_con}, \eref{Eq:Aa}, \eref{Eq:Aapprox}, and \eref{Eq:Mapprox}, we finally arrive at
\begin{eqnarray}
	\nonumber
	\fl
	\Sigma^* \approx \ln 2 + \ln \left ( 1 + \frac{\alpha ( 1 - A)}{2} \right ) - \frac{\alpha}{2} \ln \left ( \frac{4}{\alpha A} \right )
	\approx \ln 2 - \frac{\alpha}{2} 
	\left [ \ln \left ( \frac{4}{\rme \alpha A} \right ) + A \right ]\\
	\approx \ln 2 - \frac{\alpha}{2}
	\left [ \ln \left (-\frac{4\ln \alpha}{\rme \alpha } \right )
	+ \frac{1}{2} \ln \left ( \frac{\rme^2}{\pi} \ln \frac{1}{\alpha} 
	\right ) \left ( \ln \frac{1}{\alpha} \right )^{-1}\right ],
\end{eqnarray}
which is identical, up to $\Or(\alpha \ln |\ln \alpha|)$, to
$\Sigma^*$ of the AFHM given in equation (37) of \cite{Cherrier2003}.
\section{\label{App:Moment_Cal} Derivation of \eref{Eq:N_d}}
To find the moments of the number of fitness maxima $\N$, we first have
to calculate the expression defined in \eref{Eq:Sm} as
\begin{equation}
S_m \equiv  \sum_{\tsum}
	\int_{\Area(a)} \rmd  \mutvec \, p(\mutvec)  
\exp\left ( - i   \mutvec\cdot \sum_{\beta=1}^m\vk_\beta a_\beta \right ),
\end{equation}
where we introduce the short-hand notation 
\begin{equation}
\sum_{\tsum} = \sum_{a_{1}=0}^1\cdots \sum_{a_{m}=0}^1.
\end{equation}
The integral over the domain $\Area(a)$ is expressed as the difference
between the same integral over the whole space $\mathbb{R}^n$ and over
the complement $\mathbb{R}^n \setminus \Area(a)$. Accordingly, 
$S_m$ is decomposed into two parts as
\begin{equation}
S_m = 2^m F - 2^m K.
\end{equation} 
The first term simply corresponds to the characteristic function of $p(\mutvec)$, i.e.,
\begin{equation}
  \label{Eq:F_def}
F = \frac{1}{2^m} \sum_{\tsum} G\left ( \sum_{\beta=1}^m \vk_\beta 
a_\beta\right ),
\end{equation}
where $G(\vk) \equiv \int \rmd  \mutvec \, p(\mutvec) \exp\left (-\rmi \vk\cdot \mutvec\right )$.
The second term is
\begin{equation}
K = \frac{1}{2^m} \sum_{\tsum} \int_{c} \rmd  \mutvec \, p(\mutvec) \exp \left (-\rmi   \mutvec \cdot \sum_{\beta=1}^m \vk_\beta a_\beta\right ),
\end{equation}
where $\int_c$ represents the integral over the complement $\mathbb{R}^n \setminus \Area(a)$.
We can thus rewrite \eqref{Eq:Nm} as 
\begin{equation}
\langle \N^m \rangle
= \frac{2^{mL}}{(2\pi)^{mn}}
\int \prod_{\alpha=1}^m \rmd \phevec_\alpha \rmd \vk_\alpha \exp
\left [ \rmi   \vk_\alpha \cdot \phevec_\alpha- i L^\gamma \vk_\alpha \cdot \q + L \ln ( F - K) \right ], 
\label{Eq:NMomentFormal}
\end{equation}
where we have introduced the scaling relation $\Q = \q L^\gamma$ with $|\q| = \Or(1)$.

The integral \eref{Eq:NMomentFormal} can now be solved by means of the saddle point method in the limit $L\to\infty$.
Depending on the choice of the scaling of $\Q$, the integral forms a
saddle point at the scale $\vert \phevec_\alpha \vert \sim \Or(L)$ or
$\vert \phevec_\alpha \vert \sim \Or(1)$,
which determines the typical phenotypes giving rise to local maxima~\cite{Hwang2017}.
If the choice $\gamma < 1$ is made, it was shown in \cite{Hwang2017}
that typical realizations of the $\vec{\xi}_i$ can find a subset of phenotypes that are close to the origin, and thus the integral is dominated by the region 
$\vert \phevec_\alpha \vert \sim \Or(1)$ and accordingly $\vert
\vk_\alpha \vert = \Or(L^{-3/2})$.
Around this point, $F$ is expanded into
\begin{eqnarray}
\fl
F &\approx 
 \frac{1}{2^m} \sum_{\tsum} \left [ 1 - \frac{1}{2} 
\left ( \sum_{\beta=1}^m \vk_\beta
a_\beta \right ) ^2 \right ]
	= \frac{1}{2^m} \left ( 2^m - \frac{1}{2} \sum_{\alpha,\beta} \vk_\alpha \cdot \vk_\beta \sum_{\tsum}  a_\alpha a_\beta \right ) \nonumber \\
\fl
&= 1 - \frac{1}{2^{m+1}} \sum_{\alpha,\beta} \vk_\alpha \cdot \vk_\beta
\left [ \delta_{\alpha\beta} 2^{m-1} + \left ( 1 - \delta_{\alpha\beta} \right ) 2^{m-2}
\right ] 
= 1 - \sum_{\alpha,\beta} \vk_\alpha \cdot \vk_\beta A_{\alpha\beta},
\end{eqnarray}
where 
$A_{\alpha\beta} = \frac{1}{8} (1+ \delta_{\alpha\beta})$.
Note that the above approximation is valid as long as the standard
deviation of $p(\mutvec)$  is finite. In general, the sum over
$\alpha,\beta$ in the last expression is multiplied by the variance of
the distribution, which here has been set to unity. 

Next, $K$ can be expanded in a similar manner. In the region $|\phevec_\alpha| \sim \Or(1)$,
\begin{equation}
 K \approx 2^{-m} p(0) V,\quad
	V =  V[\{\phevec_\alpha\}_{\alpha = 1,\dots,m}] \equiv \sum_{\tsum} \int_{c} \rmd \mutvec 
 \sim \Or(|\phevec_\alpha|^n).
	\label{Eq:V_def}
\end{equation}
Note that the term $\rmi   \vk_\alpha \cdot \phevec_\alpha$ is negligible for this choice of $\gamma$, 
which allows the integrals over the $\vk_\alpha$'s and the $\phevec_\alpha$'s
in \eref{Eq:NMomentFormal} to be treated independently.
The integration over $\vk$'s are evaluated as follows:
\begin{eqnarray}
\int \prod_{\alpha=1}^m \rmd \vk_\alpha \exp\left ( - L \sum_{\alpha,\beta} \vk_\alpha \cdot \vk_\beta A_{\alpha\beta}
+ i L^\gamma \sum_\alpha \vk_\alpha \cdot \q \right )
\nonumber\\
	= \left (\frac{8\pi}{L}\right )^{nm/2} \frac{1}{(m+1)^{n/2}} 
\exp\left ( -L^{2\gamma-1} \frac{2m}{m+1}|\q|^2 \right ),
\label{Eq:KIntegral}
\end{eqnarray}
where we have used the fact that 
\begin{equation}
A^{-1}_{\mu\nu} = 8 \left ( \delta_{\mu\nu} - \frac{1}{m+1} \right ),
\qquad
\frac{1}{4} \sum_{\mu,\nu}\q_\mu \cdot \q_\nu A^{-1}_{\mu\nu}
= - \frac{2m}{m+1}|\q|^2.
\end{equation}
Introducing a symbol $\moment_m$ for the remaining integral over $\phevec_\alpha$, we thus obtain \eref{Eq:N_d} 
with
\begin{eqnarray}
\moment_m &=L^m \left ( \frac{2^m }{\pi^m (m+1)} \right )^{n/2} 
	\int \prod_\alpha \rmd \z_\alpha \exp\left [-Lp(0) 2^{-m} V\right ]\nonumber \\
&= \left ( \frac{2^m }{\pi^m (m+1)} \right )^{n/2}
	\int \prod_\alpha \rmd \z_\alpha  \exp\left [ -p(0) 2^{-m} V\right ],
\label{Eq:C_m}
\end{eqnarray}
where, in the last equality, we have changed the variables $L^{1/n} z_\alpha^k \mapsto z_\alpha^k$
for all components of $\z_\alpha$. 

\section{\label{Sec:Exact1d} Moments for $n=1$ and $Q = 0$}
In this appendix, we present the exact 
leading asymptotic behaviour of all moments
for the case of $n=1$ at $Q=0$.
In the following $z_\alpha$ should be
understood as a real number which can take negative values rather than the magnitude of the vector $|\phevec_\alpha|$.
Setting $n=1$ in \eref{Eq:C_m}, we write 
\begin{equation}
\moment_m = \frac{1}{\sqrt{m+1}} 
\left ( \frac{2}{\pi} \right )^{m/2} 
	\int \prod_\alpha \rmd z_\alpha \exp\left [ -p(0) 2^{-m} V\right ],
\label{Eq:tau_1d}
\end{equation}
where
\begin{eqnarray}
\fl
V &= \sum_{\tsum} \int_{c} \rmd x
= 2 \sum_{\tsum} \left [ \mathrm{max}(0,-s_1 z_1,\ldots, -s_m z_m)
- \mathrm{min}(0,-s_1 z_1,\ldots, -s_m z_m) \right ]\nonumber \\
\fl
&=4 \sum_{\tsum} \mathrm{max}(0,s_1 z_1,\ldots,s_m z_m),
\end{eqnarray}
with $s_\alpha \equiv 2 a_\alpha-1$.
In the above equation, we have used the identities
$-\mathrm{min}(0,-s_\alpha z_\alpha) = \mathrm{max}(0,s_\alpha z_\alpha)$ and
$\sum_{\tsum} \mathrm{max}(0,-s_\alpha z_\alpha) = \sum_{\tsum} \mathrm{max}(0,s_\alpha z_\alpha)$.

Since $V$ is invariant under the transformation
$z_\alpha \mapsto -z_\alpha$ for each $\alpha$ as well as 
under all permutations of the indices $\alpha$,
we can write \eref{Eq:tau_1d}, after making the change of 
variables $y_\alpha =  p(0) 2^{2-m} z_\alpha$, as
\begin{eqnarray}
\fl
\moment_m = \frac{2^{m^2}}{\sqrt{m+1}} 
\left ( \frac{2}{\pi} \right )^{m/2} \frac{1}{[4 p(0)]^m}
2^m m!\times \nonumber \\
\int_0^\infty \rmd y_1 \int_{y_1}^\infty \rmd y_2
\ldots \int_{y_{m-1}}^\infty \rmd y_m \exp \left [-\sum_{\tsum}\mathrm{max}(0,s_1 y_1,\ldots,s_m y_m)\right ].
\label{Eq:moment1D}
\end{eqnarray}
Now the domains of integration with respect to $y_\alpha$ are arranged in such a way that $y_1 < y_2 <\ldots < y_m$. 
Within this ordering, we can establish the following identity
\begin{equation}
\sum_{\tsum}\mathrm{max}(0,s_1 y
_1,\ldots,s_m y_m) = 2^{m-1} y_m + 2^{m-2} y_{m-1} + \ldots + 2 y_2 + y_1
=\sum_{k=1}^m 2^{k-1} y_k.
\end{equation}
Then, the integrals in \eref{Eq:moment1D} are computed recursively as follows:
\begin{eqnarray}
	I_m(y_{m-1}) &= \int_{y_{m-1}}^\infty \exp\left (-2^{m-1} y_m\right ) \rmd y_m = \frac{1}{2^{m-1}} \exp \left ( -2^{m-1} y_{m-1}\right ),\\
	\nonumber
I_{m-1}(y_{m-2}) &= 
	\int_{y_{m-2}}^\infty \exp \left (-2^{m-1} y_{m-1}\right ) I_m(y_{m-1}) \rmd y_{m-1} \\
	&= \frac{1}{2^{m-1}} \frac{1}{2^{m-1} + 2^{m-2}} \exp \left  [ -(2^{m-1} + 2^{m-2})y_{m-2}\right ],
\end{eqnarray}
and so on. Inserting the value $p(0) = \frac{1}{\sqrt{2 \pi}}$ for the
Gaussian distribution \eqref{Eq:common_dist}, we thus get
\begin{equation}
\moment_m = \frac{2^{m^2} m!}{\sqrt{m+1}} \prod_{k=1}^m \frac{1}{\sum_{j=1}^k 2^{m-j}} 
= \frac{2^{m^2} m!}{\sqrt{m+1}} \prod_{k=1}^m \frac{1}{2^{m} - 2^{m-k}}= \frac{m!}{\sqrt{m+1}} \prod_{k=1}^m \frac{1}{1-2^{-k}}.
\label{Eq:moment}
\end{equation}
The first few moments are $\moment_1 = \sqrt{2}$, $\moment_2 =
16/\sqrt{27}$ and $\moment_3 = 64/7$. For general distributions
$p(\xi)$ with zero mean and variance $\omega^2$ the expression
\eqref{Eq:moment} is multiplied by a factor according to
\begin{equation}
  \label{Eq:mom_gen}
  \moment_m \mapsto \left( 2 \pi \omega^2 p(0)^2 \right)^{-\frac{m}{2}} \moment_m.
  \end{equation}

\section{\label{Sec:QPoch}The $q$-Pochhammer symbol}
This appendix summarises some properties of the $q$-Pochhammer symbol that 
are used in this paper. The $q$-Pochhammer symbol was defined in \eqref{Eq:Pochhammer}.
From the definition, we obtain
\begin{equation}
(a;q)_k = (-1)^k a^k q^{k(k-1)/2} \left (a^{-1};q^{-1} \right )_k.
\label{Eq:Qin}
\end{equation}
If $(a;q)_\infty$ exists, we can write
\begin{equation}
\frac{1}{(a;q)_k} = \frac{1}{(a;q)_\infty} \prod_{l=k}^\infty (1 - a q^l)
= \frac{(aq^k;q)_\infty}{(a;q)_\infty}
\label{Eq:Qinf}
\end{equation}
Using \eref{Eq:Qinf} and the infinite series representation 
\begin{equation}
(qx;q)_\infty 
	= \sum_{l=0}^\infty \frac{x^l}{(q^{-1};q^{-1})_l},
\label{Eq:Qid}
\end{equation}
we can write for $q=\frac{1}{2}$
\begin{equation}
\frac{1}{(\frac{1}{2};\frac{1}{2})_k} = S \sum_{l=0}^\infty \frac{2^{-kl}}{(2;2)_l},
\label{Eq:Qcon}
\end{equation}
where $S \equiv \left [ (\frac{1}{2};\frac{1}{2})_\infty \right ]^{-1}\approx 3.462\,7466$.

Let 
\begin{equation}
a_k \equiv S \sum_{l=0}^\infty \frac{l^k 2^l}{(2;2)_l}
= S \left . \left ( x \frac{\rmd}{\rmd x} \right )^k  \left ( \frac{x}{2} ; \frac{1}{2}
\right )_\infty \right |_{x=2},
\end{equation}
where we have used \eref{Eq:Qid} to obtain the differential form.
As the sum converges quickly, the partial sum of the first few terms already produces an accurate estimate
of $a_k$. The error of the $l_0$th order approximation is given by
\begin{equation}
e_k \equiv S \left | \sum_{l=l_0+1}^\infty \frac{l^k 2^l }{(2;2)_l} \right |
\le S \sum_{l=l_0+1}^\infty \frac{l^k 2^l}{| (2;2)_l|}.
\end{equation}
Since
\begin{equation}
\left | \frac{1}{(2;2)_l}\right | 
= 2^{-l(l+1)/2} \prod_{k=1}^l (1 - 2^{-k})^{-1} \le S 2^{-l(l+1)/2} ,
\end{equation}
we have
\begin{eqnarray}
e_k &\le S^2 \sum_{l=l_0+1}^\infty l^n 2^{-l(l-1)/2}  
= S^2 l_0^n \sum_{k=1}^\infty \left (1+\frac{k}{l_0} \right )^k 
2^{-(l_0+k)(l_0+k-1)/2}\nonumber \\
&\le S^2  l_0^k 2^{-l_0(l_0-1)/2} 
\sum_{r=1}^\infty
\exp\left [ - \frac{\ln 2}{2} r^2 - \left ( \frac{2 l_0 -1}{2} \ln 2 - \frac{k}{l_0} \right ) r \right ],
\end{eqnarray}
where we use that $1+x \le \rme^x$ for $x \ge 0$.
If we choose $l_0$ such that $l_0 (2 l_0 - 1) \ln 2 - 2 k \ge 0$, 
we get
\begin{eqnarray}
\fl
e_k &\le S^2 l_0^k 2^{-(l_0^2 + l_0 + 1)/2} \rme^{k/l_0}
	\sum_{r=1}^\infty \exp \left ( - \frac{\ln 2}{2}r^2 \right )\nonumber \\
\fl
	&\le S^2 l_0^k 2^{-(l_0^2 + l_0 + 1)/2} \rme^{k/l_0} \int_0^\infty \rmd r \exp\left (-\frac{\ln 2}{2} r^2\right )
= S^2 \sqrt{\frac{\pi}{\ln 4}} l_0^k 2^{-(l_0^2 + l_0 + 1)/2}\rme^{k/l_0}.
\end{eqnarray}
For example, if we choose $l_0 = 12$ for $k = 5$, 
we obtain $e_5 \le 1.6\times 10^{-17}$.

In particular, we can get exact formulae for $k=0$ and $k=1$. 
Since
\begin{equation}
\sum_{l=0}^\infty \frac{2^l}{(2;2)_l} = \left (1;\frac{1}{2} \right )_\infty =0,
\end{equation}
we trivially have $a_0 = 0$.
To find $a_1$, 
we write
\begin{equation}
\left (\frac{x}{2};\frac{1}{2} \right )_\infty 
\equiv \left ( 1 - \frac{x}{2} \right )
g(x),
\end{equation}
where 
\begin{equation}
g(x) = \prod_{l=1}^\infty \left (1 - \frac{x}{2^{l+1}} \right ).
\end{equation}
Note that $g(2) = S^{-1}$.
From this identity, we find
\begin{equation}
a_1 = \frac{x}{g(2)} \left .  \frac{\rmd }{\rmd x} \left [ \left ( 1 - \frac{x}{2} 
\right ) g(x) \right ] \right |_{x=2} = -1.
\end{equation}

\section{\label{App:Another}Another way of finding $P(x)$}
We first observe that
\begin{equation}
\frac{1}{\sqrt{m+1}} = \frac{1}{\sqrt{\pi}} \int_0^\infty \frac{\rme^{-(m+1)t}}{\sqrt{t}} \rmd      t.
\end{equation}
Inserting this into \eqref{Eq:GkSer} yields
\begin{eqnarray}
\G(k) &= \frac{1}{\sqrt{\pi}} \sum_{m=0}^\infty \frac{(\rmi k)^m}{(\frac{1}{2};\frac{1}{2})_m}  \int_0^\infty
\frac{\rme^{-(m+1)t} }{\sqrt{t}} \rmd      t
= \frac{1}{\sqrt{\pi}} \int_0^\infty \rmd      t \frac{\rme^{-t}}{\sqrt{t}}
\sum_{m=0}^\infty \frac{(\rmi k\rme^{-t})^m}{(\frac{1}{2};\frac{1}{2})_m} \nonumber \\
&= \frac{1}{\sqrt{\pi}} \int_0^\infty \rmd      t \frac{\rme^{-t}}{\sqrt{t}} \prod_{l=0}^\infty \frac{1}{1 - \rmi k \rme^{-t} 2^{-l}},
\end{eqnarray}
where we have exchanged the orders of summation and integration to arrive
at the second equality and used the relation 
\begin{equation}
\sum_{m=0}^\infty \frac{x^m}{(q;q)_m}
= (x;q)_\infty^{-1} 
\end{equation}
to obtain the last equality.
Hence
\begin{eqnarray}
	P(x) &= \frac{1}{2\pi} \int_{-\infty}^\infty \rmd k \exp\left (-\rmi kx\right ) \G(k) \nonumber \\
	&= \frac{1}{\sqrt{\pi}} \int_0^\infty \rmd      t \frac{\rme^{-t}}{\sqrt{t}}
	\int \rmd k \frac{\exp\left (-\rmi kx\right )}{2\pi}  \prod_{l=0}^\infty \frac{1}{1 - \rmi k \rme^{-t} 2^{-l}},
\end{eqnarray}
where we again changed the order of integration.
Since there are poles at $k = -\rmi \rme^t 2^l$ ($l=0,1,2,\ldots$) 
in the complex $k$ plane, 
$P(x) = 0$ for $x<0$. The integral over $k$ for $x>0$ can be performed as
\begin{eqnarray}
\fl
	\frac{1}{2\pi} \int \rmd k \exp\left (-\rmi kx\right ) \prod_{l=0}^\infty \frac{1}{1 - \rmi k \rme^{-t} 2^{-l}}
= \rme^t \sum_{m=0}^\infty 2^m \exp\left (-2^m x \rme^t\right ) \prod_{l\neq m} \frac{1}{1 - 2^{m-l}}
\nonumber \\
\fl
=S \rme^t \sum_{m=0}^\infty 2^m \exp \left ( -2^m x \rme^t \right ) \prod_{l=1}^{m} \frac{1}{1-2^l} 
= S \rme^t \sum_{m=0}^\infty 2^m \exp \left ( -2^m x \rme^t \right ) \frac{1}{(2;2)_m},
\end{eqnarray}
which gives
\begin{eqnarray}
\fl
P(x) &= \frac{S}{\sqrt{\pi}} \sum_{m=0}^\infty \frac{2^m}{(2;2)_m}\int_0^\infty \rmd      t 
\frac{\exp \left ( -2^m x \rme^t\right )}{\sqrt{t}}
= \frac{S}{\sqrt{\pi}} \sum_{m=0}^\infty \frac{2^m}{(2;2)_m}\int_1^\infty \rmd      t 
\frac{\exp\left (-2^m x t\right )}{t \sqrt{\ln t}} \nonumber \\
\fl
&\equiv S \sum_{m=0}^\infty \frac{2^m}{(2;2)_m} \psi(2^m x),
\label{Eq:prob}
\end{eqnarray}
where
\begin{equation}
\psi(x) \equiv 
\frac{1}{\sqrt{\pi}} \int_1^\infty \frac{\rme^{-xt}}{t\sqrt{\ln t}} \rmd      t
	=\int_0^\infty \frac{\exp \left (-x\rme^t\right )}{\sqrt{\pi t}} \rmd      t
      \end{equation}
      in agreement with \eqref{Eq:psix}.
To confirm, we calculate the $m$th moment $\moment_m$ from \eref{Eq:prob} as
\begin{eqnarray}
	\fl
\moment_m &\equiv \int_0^\infty x^m P(x) \rmd x
= \frac{S}{\sqrt{\pi}} \sum_{p=0}^\infty \frac{2^p}{(2;2)_p}\int_0^\infty 
\frac{\rmd t}{\sqrt{t}} \int_0^\infty x^m \exp\left (-2^p \rme^t x\right ) \rmd x\nonumber \\
\fl
&=  m!\frac{S}{\sqrt{\pi}} \sum_{p=0}^\infty \frac{2^p}{(2;2)_p}\int_0^\infty 
\frac{\rmd t}{\sqrt{t}} 2^{-p(m+1)} \rme^{-(m+1)t}\nonumber \\
\fl
&= \frac{m!}{\sqrt{m+1}} S \sum_{p=0}^\infty \frac{2^{-mp}}{(2;2)_p} = 
\frac{m!}{\sqrt{m+1}} \frac{(2^{-m-1};\frac{1}{2} )_\infty }{(\frac{1}{2};\frac{1}{2})_\infty}
=\frac{m!}{\sqrt{m+1} (\frac{1}{2};\frac{1}{2})_m},
\end{eqnarray}
which is the desired result.

\section{\label{App:Asym}Asymptotic behaviour of $P(x)$}
When $x\gg 1$, we can approximate \eqref{Eq:psix} as
\begin{eqnarray}
\fl
\psi(x) &= \frac{1}{\sqrt{\pi}} \rme^{-x}
\int_0^\infty \frac{\rme^{-x t}}{(1+t)\sqrt{\ln (1+t)}} \rmd      t
\approx \frac{1}{\sqrt{\pi}} \rme^{-x} \left [ \int_0^\infty \frac{\rme^{-xt}}{\sqrt{t}} \rmd      t + \Or(\rme^{-x}) \right ] \nonumber \\
\fl
&= \frac{\rme^{-x}}{\sqrt{x}} +\Or(\rme^{-2x})
\end{eqnarray}
Since the terms with $l\ge 1$ in \eqref{Eq:PN} contribute at most $\Or(\rme^{-2x})$, the leading behaviour
of $P(x)$ is $S \rme^{-x}/\sqrt{x}$.

For small $x$, we write $\psi(x) = (I_1 + I_2 + I_3)/\sqrt{\pi}$
with 
\begin{eqnarray}
	I_1 = \int_0^\mlnx \frac{\exp \left (-\rme^{t-\mlnx}\right )}{\sqrt{t}} \rmd      t
	= \int_0^\mlnx \frac{\exp\left (-\rme^{-t}\right )}{\sqrt{\mlnx-t}} \rmd t,\nonumber \\
	I_2 = \int_0^\mlnx \frac{\exp\left ( - \rme^t\right )}{\sqrt{t+\mlnx}} \rmd      t,\quad
	I_3 =  \int_\mlnx^\infty \frac{\exp\left ( - \rme^t\right )}{\sqrt{t+\mlnx}} \rmd      t,
\end{eqnarray}
and $\mlnx = -\ln x$.
$I_3$ is at most $\Or(\rme^{-\mlnx})$ because
\begin{equation}
I_3 =\int_0^\infty \frac{\exp(-\rme^{t+\mlnx})}{\sqrt{t+2\mlnx}} \rmd      t
\le  \frac{1}{\sqrt{2 \mlnx}} \int_0^\infty \exp(-t\rme^\mlnx) \rmd      t
= \frac{\rme^{-\mlnx}}{\sqrt{2 \mlnx}}.
\end{equation}
Next, we find the asymptotic behaviour of $I_2$ as
\begin{eqnarray}
\fl
I_2  
&= \frac{1}{\sqrt{\mlnx}} \int_0^\mlnx \exp\left (-\rme^{t}\right ) \left ( 1 + \frac{t}{\mlnx}\right )^{-1/2} \rmd      t  \nonumber 
= \frac{1}{\sqrt{\mlnx}} \int_0^\mlnx \exp(-\rme^{t}) \left [ 1 - \frac{t}{2 \mlnx} +\Or(\mlnx^{-2})\right ] \rmd      t \nonumber \\
\fl
&= \frac{1}{\sqrt{\mlnx}}
\left [ \lambda_1 -\frac{\lambda_2}{\mlnx} + \Or(\mlnx^{-2}) \right ],
\end{eqnarray}
where 
\begin{equation}
\lambda_1 = \int_0^\infty \exp\left (-\rme^t\right ) \rmd      t, \qquad 
\lambda_2 = \frac{1}{2}\int_0^\infty t \exp\left (-\rme^t\right ) \rmd      t,
\end{equation}
and we have used 
\begin{equation}
\int_\mlnx^\infty  \frac{t^n}{\exp\left (\rme^t\right )} \rmd      t 
=\int_0^\infty  \frac{(\mlnx+t)^n}{\exp(\rme^{\mlnx+t})} \rmd      t 
	\le \mlnx^n \int_0^\infty \exp\left ( -\left (\rme^\mlnx - \frac{n}{\mlnx}\right ) t\right )\rmd      t = \Or(\rme^{-\mlnx}).
\end{equation}
The leading behaviour of $\psi$ comes from $I_1$,
\begin{eqnarray}
\fl
I_1 = \left . 2 \sqrt{\mlnx-t} \exp(-\rme^{-t}) \right |_{t=0}^\mlnx 
+ 2 \int_0^\mlnx \sqrt{\mlnx-t} \exp(-t -\rme^{-t}) \rmd      t \nonumber \\
\fl
=\frac{2}{e} \sqrt{\mlnx} + 2 \sqrt{\mlnx} \int_0^\mlnx \exp(-t -\rme^{-t}) (1 - t/\mlnx)^{1/2} \rmd      t\nonumber \\
\fl
\approx  \frac{2}{e} \sqrt{\mlnx} + 2 \sqrt{\mlnx}
 \int_0^\mlnx \exp(-t -\rme^{-t}) \left ( 1 - \frac{t}{2\mlnx} -\frac{t^2}{8\mlnx^2}
\right ) \rmd      t \nonumber \\
\fl
= 
2 \sqrt{\mlnx}
- 2 \sqrt{\mlnx}
\int_0^\infty \exp(-t -\rme^{-t}) \left ( \frac{t}{2\mlnx} +\frac{t^2}{8\mlnx^2} \right ) \rmd      t 
= 2 \sqrt{\mlnx} - \frac{\lambda_3}{\sqrt{\mlnx}} - \frac{\lambda_4}{\mlnx^{3/2}},
\end{eqnarray}
with
\begin{eqnarray}
\fl
	\lambda_3 &= \int_0^\infty t \exp\left (-t -\rme^{-t}\right ) \rmd      t 
	= \int_{-\infty}^\infty t\exp \left ( -t-\rme^{-t}\right ) \rmd      t 
	- \int_{-\infty}^0 t \exp\left (-t -\rme^{-t}\right ) \rmd      t,\nonumber \\
\fl
	&= \gamma + \int_0^\infty t \exp\left (t-\rme^t\right ) \rmd      t
	= \gamma -\left . t \exp \left (-\rme^t\right ) \right |_0^\infty + \int_0^\infty \exp \left (-\rme^{t}\right ) \rmd      t
= \gamma + \lambda_1,\\
\fl
	\lambda_4 &= \int_0^\infty \frac{t^2}{4} \exp \left (-t -\rme^{-t}\right ) \rmd      t
	= \int_{-\infty}^\infty \frac{t^2}{4} \exp\left (-t -\rme^{-t}\right ) \rmd      t
	- \int_{-\infty}^0 \frac{t^2}{4}\exp \left (-t - \rme^{-t}\right ) \rmd      t\nonumber \\
\fl
	&=\frac{\gamma^2}{4} + \frac{\pi^2}{24} - \int_{0}^\infty \frac{t^2}{4}   \exp\left (t-\rme^t\right ) \rmd      t 
	\nonumber\\
	\fl
	&=\frac{\gamma^2}{4} + \frac{\pi^2}{24}+ \left . \frac{t^2}{4} \exp\left ( -\rme^t \right ) \right |_{t=0}^\infty - \frac{1}{2} \int_{0}^\infty t  \exp\left (-\rme^t\right ) \rmd      t
=\frac{\gamma^2}{4} + \frac{\pi^2}{24} - \lambda_2,
\end{eqnarray}
where $\gamma \approx 0.5772$ is the Euler-Mascheroni number.
Hence the asymptotic behaviour of $\psi(x)$ is 
\begin{equation}
\psi(x) = \frac{2}{\sqrt{\pi}} \sqrt{\mlnx} - \frac{\gamma}{\sqrt{\pi \mlnx}}
- \left (\frac{\gamma^2}{4} + \frac{\pi^2}{24} \right ) \frac{1}{ \sqrt{\pi}\mlnx^{3/2}}
= \frac{2}{\sqrt{\pi}} \sqrt{\mlnx} - \frac{\gamma}{\sqrt{\pi \mlnx}}
- \frac{\lambda}{\sqrt{\pi}\mlnx^{3/2}},
\end{equation} 
where $\lambda \approx 0.494~528$. 
Since 
\begin{eqnarray}
\fl
	\sqrt{\pi} \psi(2^l x)   &=2\left (\mlnx-l \ln 2\right )^{1/2} - \gamma(\mlnx-l\ln 2)^{-1/2} \nonumber 
- \lambda(\mlnx - l \ln 2)^{-3/2}\\
\fl
 &=2\sqrt{\mlnx}-\frac{l \ln 2 + \gamma }{\sqrt{\mlnx }} 
+\mlnx^{-3/2}
    \left(-\lambda -\frac{1}{4} l^2 \ln ^2 2-\frac{1}{2} \gamma  l \ln
     2\right)\nonumber \\
\fl
&\quad-\frac{1}{8} \mlnx^{-5/2} \left(l^3 \ln^3 2+3 \gamma  l^2 \ln ^2 2+12 \lambda  l \ln  2\right),
\end{eqnarray}
we get
\begin{eqnarray}
\fl
P(x) &\approx  \frac{\ln 2}{\sqrt{\pi}}  \mlnx^{-1/2} - 
\frac{a_2 \ln^2 2 - 2 \gamma \ln 2}{4\sqrt{\pi}}\mlnx^{-3/2}
+ \frac{12 \lambda \ln 2 - 3 \gamma a_2 \ln^2 2 - a_3 \ln^3 2}{8\sqrt{\pi}} \mlnx^{-5/2}\nonumber \\
\fl
&= \frac{\ln 2}{\sqrt{-\pi \ln x}}
\left ( 1 + \frac{0.094\,944}{\ln x} + 
\frac{0.150\,994}{\ln^2 x}
\right ).
\end{eqnarray}

\section{\label{Sec:Sub}Finite size corrections to $\langle \N^m
  \rangle$ for $n=1$}
In this appendix, we compute the finite size corrections to
\eref{Eq:N_d} for $n=1$ and $q = 0$. 
To this end, we  expand $F$ defined in \eqref{Eq:F_def} up to fourth order of $k$,
\begin{equation}
	F \approx 1 - P+B,\quad 
	P\equiv \sum_{\alpha,\beta} k_\alpha A_{\alpha\beta} k_\beta,
	\quad B\equiv \frac{1}{2^{m}} \frac{1}{8} \sum_{\tsum}\left(
				\sum_{\alpha} k_\alpha a_\alpha
			\right)^4 .
\end{equation}
The quantity $B$ can be expressed as
\begin{eqnarray}
\fl
	2^{m+3} B &= \sum_{j_1} k_{j_1}^4 2^{m-1} + \sum_{j_1 \ne j_2} k_{j_1}^3 k_{j_2} 2^{m-2} \binom{4}{3}
+  \sum_{j_1 < j_2} k_{j_1}^2 k_{j_2}^2 2^{m-2} \binom{4}{2}  \nonumber \\
\fl
	&+ \sum_{j_1} \sum_{  j_2 < j_3 , j_1 \ne j_2,j_3} k_{j_1}^2 k_{j_2}k_{j_3} 2^{m-3} \binom{4}{2} 2 
	+ \sum_{j_1 < j_2 < j_3 < j_4} k_{j_1} k_{j_2}k_{j_3} k_{j_4}2^{m-4} 4!. 
\end{eqnarray}
Expanding the higher orders up to $ \Or(k_i^4) $ and $ \Or(z_i^2) $,
we have for large $L$
\begin{eqnarray}
\fl
	\Avr{\N^m} &= \frac{2^{mL}}{(2\pi)^m} \int \prod_{i=1}^{m} \rmd z_i \rmd k_i \exp \left[ L \ln \left(
			1 - P + B - K
		\right)		
	\right] \nonumber \\
\fl
	&= \frac{2^{mL}}{(2\pi)^m} \int \prod_{i=1}^{m} \rmd z_i \rmd k_i 
	\exp \left[ - L (P  + K ) \right] 
	 \left(
		1 + B L-\frac{L K^2}{2}-L K P-\frac{L P^2}{2}
   \right).
   \label{Eq:Mom_fin}
\end{eqnarray}
The terms in the parenthesis are simply a collection of multi-variate polynomials of $k_i$'s.
They are evaluated on a case-by-case basis 
using the following formulae:
\begin{eqnarray}
	\int \prod_{i=1}^{m} \rmd k_i \rme^{-LP} k_j^4 
	= \frac{48 m^2}{L^2 (m+1)^2} \frac{1}{L^{m/2}} \sqrt{\frac{(8\pi)^m}{m+1}},\nonumber \\ 
	\int \prod_{i=1}^{m} \rmd k_i \rme^{-LP} k_{j_1}^3 k_{j_2} 
	= -\frac{48 m}{L^2 (m+1)^2} \frac{1}{L^{m/2}} \sqrt{\frac{(8\pi)^m}{m+1}},\nonumber \\ 	
	\int \prod_{i=1}^{m} \rmd k_i \rme^{-LP} k_{j_1}^2 k_{j_2}^2 
	= \frac{16 \left(m^2+2\right)}{L^2 (m+1)^2} \frac{1}{L^{m/2}} \sqrt{\frac{(8\pi)^m}{m+1}},\nonumber \\ 		
	\int \prod_{i=1}^{m} \rmd k_i \rme^{-LP} k_{j_1}^2 k_{j_2} k_{j_3}
	= -\frac{16 (m-2)}{L^2 (m+1)^2} \frac{1}{L^{m/2}} \sqrt{\frac{(8\pi)^m}{m+1}},\nonumber \\ 			
	\int \prod_{i=1}^{m} \rmd k_i \rme^{-LP} k_{j_1} k_{j_2} k_{j_3} k_{j_4}
	= \frac{48}{L^2 (m+1)^2} \frac{1}{L^{m/2}} \sqrt{\frac{(8\pi)^m}{m+1}}, 
\end{eqnarray}
where the indices of $k$ in the integrals on the left-hand side are
assumed to be different.
Integrating out the $k_i$ in \eqref{Eq:Mom_fin}, we get
\begin{eqnarray}
	\fl
	\Avr{\N^m} 
		= &\frac{2^{mL}}{(2\pi)^m} \frac{1}{L^{m/2}} \sqrt{\frac{(8\pi)^m}{m+1}} \int \prod_{i=1}^{m} \rmd z_i \exp\left ( - L K \right )\nonumber \\
		\fl
	&\times  
		\left[
				1 + \frac{3\ m \left(m^2+m+2\right)}{8 L (m+1)}-\frac{L K^2}{2}- K \frac{m}{2} - \frac{m (m+2)}{8 L}
		\right].
\end{eqnarray}
Finally, using \eref{Eq:moment}
\begin{equation}
	\int \prod_{i=1}^{m} \rmd z_i \exp \left(
						- L K 
			\right) = \frac{m!}{L^m} \left(
				\frac{\pi}{2}
			\right)^{m/2} Q_m
\end{equation}
with $Q_m = \left [ (\frac{1}{2};\frac{1}{2})_m\right ]^{-1}$, we have 
\begin{equation}
	\Avr{\N^m} 
	= Q_m  \frac{m! 2^{L m} L^{-3 m/2}}{\sqrt{m+1}} \left[
				1 -\frac{3 m^2 (m+2)}{4 L (m+1)}
		\right].
\end{equation}

\section{\label{Sec:Simul} Numerical estimate of $P(\scaledN)$ for large $L$}
The probability density $P(\scaledN)$ of the
rescaled random variable \eqref{Eq:scaledN}  can be computed by
counting the number of local maxima for many different fitness landscape realizations. We will refer to this algorithm as the exact enumeration (EE)
method.
Since the number of genotypes increases exponentially
with $L$, the EE method becomes unfeasible for sufficiently
large $L$. 
To circumvent this difficulty, 
we employ a trick to count the
number of local maximum for a given fitness landscape. 
This appendix explains our numerical method used for $n=1$, 
but the extension to higher dimensions is straightforward.

Since the number of local maxima is on average $\sim 2^LL^{-3/2}$, the probability of a randomly
chosen genotype being a local maximum is $\sim L^{-3/2}$. For a given fitness landscape, 
we choose $M$ genotypes randomly and check if the chosen genotype is
a local maximum. If there are $m$ local maxima out of $M$ randomly chosen
genotypes, we evaluate $\scaledN$ as
\begin{equation}
\scaledN \approx \frac{m}{M} L^{3/2} ,
\end{equation}
because $\N/2^{L}$ is the probability that a randomly chosen genotype is a local maximum.

We choose $M$ such that the bin size is larger than the expected statistical
error of the Monte Carlo method. 
With 99\% probability, $m$ should lie in the interval 
\begin{equation}
\left | m - M p \right | < 3 \sqrt{Mp},
\end{equation}
where $ p = \scaledN L^{-3/2}$.
Accordingly,
\begin{equation}
\left | \frac{m}{M} L^{3/2} - \scaledN \right | < 3 \sqrt{\frac{\scaledN L^{3/2}}{M}}.
\end{equation}
Notice that for $M = 10^5 L^{3/2}$ the statistical error is about 0.01 when $\scaledN\approx 1$. 
Thus, in simulations, we set $M=10^5 L^{3/2}$ and choose the bin size 0.01.

When $M$ is smaller than $2^L$, this Monte Carlo approach is more
efficient than the EE method. 
As a rule of thumb, the Monte Carlo method is found to be more efficient than the EE method if $L \ge 24$.

\section{\label{Sec:AFH} Anti-ferromagnetic Hopfield model for finite $n$}
In this section, we present an analytic expression for the moments of
the number $\N$ of local energy minima of
the AFHM for finite $n$ and derive the full distribution for $n=1$. To
exploit the similarity to FGM we rewrite the Hamiltonian \eqref{Eq:H_AFM} in
the form
\begin{equation}
	H_\mathrm{AFHM}(\genotype) 
	=\frac{1}{4L} \vert \phevec(\genotype) \vert^2,\quad
	\phevec(\genotype)\equiv \sum_i \mutvec_i s_i(\genotype) ,
\end{equation}
where $\genotype$ now denotes a configuration of Ising spins $s_i \pm 1$
and the $\mutvec_i$'s are i.i.d.~random variables with a joint
distribution $p(\mutvec)$.
Since the calculations are largely analogous to those for FGM, we just sketch the procedure
and present the results.

The condition for a spin configuration $\genotype_\alpha$ to be a local maximum is
($i=1,2,\ldots,L$)
\begin{equation}
	\left |\phevec_\alpha -2s_i \mutvec_i \right | =
	\left |2\mutvec_i-s_i\phevec_\alpha \right | 
	> \left | \phevec_\alpha \right |,
\end{equation}
where $\phevec_\alpha \equiv \phevec(\genotype_\alpha)$. Thus
the condition for $m$ configurations to be simultaneous local minima
can be written as
\begin{equation}
\mutvec_i \in \AArea_i \equiv \bigcap_{\alpha=1}^m
	\D\left [ \frac{1}{2} s_{i,\alpha} \phevec_\alpha\right ],
\end{equation}
where $\D$ is defined in \eref{Eq:area_def}.
By replacing $\tau_i \mapsto s_i$ and $\Area \mapsto \AArea$
in the calculations for FGM, 
it is straightforward to find the $m$th moment of $\N$, which is
given by
\begin{equation}
\langle \N^m \rangle
	=\int_{\mathbb{R}^n} \prod_{\alpha=1}^m \frac{\rmd  \phevec_\alpha \rmd  \vk_\alpha}{(2\pi)^{n}} \exp\left (\rmi \vk_\alpha \cdot \phevec_\alpha\right ) \left ( S_m\right )^L
\end{equation}
with
\begin{equation}
S_m \equiv  \sum_{\tsum}
	\int_{\AArea(a)} \rmd  \mutvec \, p(\mutvec)  
\exp\left ( - i   \mutvec\cdot \sum_{\beta=1}^m\vk_\beta a_\beta \right ),
\quad
\sum_{\tsum} = \sum_{a_{1}=-1}^1\cdots \sum_{a_{m}=-1}^1,
\label{Eq:ASm}
\end{equation}
and  the domain of integration 
\begin{equation}
	\AArea(a) \equiv \bigcap_{\alpha=1}^m \D \left [  \frac{1}{2} a_\alpha \phevec_\alpha \right ].
\end{equation}
Note that $a_\alpha$ now takes the values $\pm 1$.

The calculation of $S_m$ is almost identical to that in \ref{App:Moment_Cal}.
Decomposing $S_m$ into two parts
\begin{eqnarray}
	\fl
	\nonumber
S_m = 2^m \widetilde F - 2^m \widetilde K,\\
\fl
\widetilde F = \frac{1}{2^m} \sum_{\tsum} G\left ( \sum_{\beta=1}^m \vk_\beta 
a_\beta\right ),\quad
\widetilde K = \frac{1}{2^m} \sum_{\tsum} \int_{c} \rmd  \mutvec \, p(\mutvec) \exp \left (-\rmi   \mutvec \cdot \sum_{\beta=1}^m \vk_\beta a_\beta\right ),
\end{eqnarray}
where $G(\vk)$ is the Fourier transform of $p(\mutvec)$ and 
$\int_c$ represents the integral over the complement $\mathbb{R}^n \setminus \AArea(a)$,
we can write
\begin{equation}
\langle \N^m \rangle
= \frac{2^{mL}}{(2\pi)^{mn}}
\int \prod_{\alpha=1}^m \rmd \phevec_\alpha \rmd \vk_\alpha \exp
\left [ \rmi   \vk_\alpha \cdot \phevec_\alpha+ L \ln ( \widetilde F - \widetilde K) \right ].
\label{Eq:ANMomentFormal}
\end{equation}
Repeating the same procedure as in \ref{App:Moment_Cal}, we get 
\begin{eqnarray}
\widetilde F \approx 
 \frac{1}{2^m} \sum_{\tsum} \left [ 1 - \frac{1}{2} 
\left ( \sum_{\beta=1}^m \vk_\beta
a_\beta \right ) ^2 \right ]
	= \frac{1}{2^m} \left ( 2^m - \frac{1}{2} \sum_{\alpha,\beta} \vk_\alpha \cdot \vk_\beta \sum_{\tsum}  a_\alpha a_\beta \right ) \nonumber \\
= 1 - \frac{1}{2^{m+1}} \sum_{\alpha,\beta} \vk_\alpha \cdot \vk_\beta
 \delta_{\alpha\beta} 2^{m}
	= 1 - \frac{1}{2}\sum_{\alpha} \vk_\alpha^2 ,\nonumber\\
 \widetilde K \approx 2^{-m} p(0) \widetilde V,\quad
\widetilde V \equiv \sum_{\tsum} \int_{c} \rmd \mutvec 
 \sim \Or(|\phevec_\alpha|^n).
\end{eqnarray}
Note that $\widetilde V = V(\phevec/2)$ 
and, accordingly, $\widetilde K \approx K 2^{-nm}$,
where $V$ and $K$ are defined in \eref{Eq:V_def}. 
Integration over the $\vk_\alpha$'s followed by the integration over the $\phevec_\alpha$'s gives
\begin{equation}
  \label{Eq:Moments_AFHM}
	\langle \N^m \rangle \approx 
	\left [ \frac{2^L}{(2 \pi L)^{n/2}} \right ]^m \int \prod_\alpha \rmd \phevec_\alpha e^{-L\widetilde K}
	=\left ( \frac{2^{L}}{ L^{1+n/2}} \right )^m 
	\left ( m+1\right )^{n/2} \mu_m,
\end{equation}
where we have changed the variables $L^{1/n} z_\alpha^k/2 
\mapsto z_\alpha^k$ and $\mu_m$ is defined in \eref{Eq:C_m}.

Since the explicit form of $\mu_m$ for $n=1$ is known, the moments for
the one-dimensional AFHM are given by  
\begin{equation}
  \label{Eq:Mom_AFHM}
	\langle \N^m \rangle \approx 
	\left ( \frac{2^L}{L^{3/2}}\right )^m Q_m m!,
\end{equation}
where $Q_m$ is defined in \eref{Eq:1dtau}. Defining again the rescaled
random variable $X$ through \eqref{Eq:X1D} for $L \to \infty$, 
we can write down its generating function as 
\begin{eqnarray}
\G(k) &= 
 \sum_{m=0}^\infty Q_m(\rmi k)^m=
	S \sum_{m=0}^\infty (\rmi k)^m
\sum_{l=0}^\infty \frac{2^{-lm}}{(2;2)_l} \\
&= S\sum_{l=0}^\infty \frac{1}{(2;2)_l} \sum_{m=0}^\infty (\rmi k2^{-l})^m
	= S \sum_{l=0}^\infty \frac{1}{(2;2)_l}\frac{1}{1-\rmi k 2^{-l}},\nonumber 
\label{Eq:AG1}
\end{eqnarray}
where the analytic continuation has been easily attained.
It is now straightforward to find the probability density $P(x)$, which is
given by \eqref{Eq:P_AFHM} in the main text. 

Next we discuss the asymptotic behaviour of $P(x)$.
For large $x$ the term with $l=0$ dominates,
which gives $P(x) \sim S e^{-x}$.
To find the asymptotics for small $x$, we first
note that $P(0)=S(1;\frac{1}{2})_\infty=0$; see \eref{Eq:Qid}.
In fact, the $k$th derivative of $P(x)$ at $x=0$ is
$P^{(k)}(0) = (-1)^k S (2^{k-1};\frac{1}{2})_\infty = 0$, so
$P(x)$ near $x=0$ is hardly discernible from 0.
For small $0<x \ll 1$, the dominant contribution is expected when $2^l x\le 1$, or $l \le -\ln x/\ln 2\equiv
l_x$.
By approximating $\exp(-2^l x) \approx \theta ( l_x - l)$,
we have
\begin{equation}
	P(x) \approx \left | -S \sum_{l=l_x}^\infty \frac{2^l}{(2;2)_l} \right |
	\approx S 2^{-l_x(l_x+3)/2} = \exp(-a (\ln x)^2 + b \ln x +c),
\end{equation}
where we use that $\sum_{l=0}^\infty \frac{2^l }{(2;2)_l} = 0$, $2^{-l_x} = x$, and we neglect the sign because
$P(x)$ should be positive. Although we cannot find an analytic form of
the parameters $a$ and $b$,
fitting gives a reasonable result with $a = 0.851$, $b =1.64$, $c=-0.215$.
\Fref{Fig:AFN} shows the probability density in comparison to the asymptotic behaviour.

As a minimal check of the validity of these results,
we calculated a few moments using Monte Carlo simulations along the
lines of \ref{Sec:Simul}.  
Note that $ \langle (X L^{-3/2})^m \rangle$ is the probability that
$m$ randomly chosen configurations
are all local minima for a random Hamiltonian.
To calculate moments, we first generate $L$ random variables
$\mutvecM_i$, and then
choose one set of $m$ random configurations, to check if these configurations
are all local minima. If $e$ sets of configurations are found to be local mimima among $E$ such attempts
(that is, $E$ random Hamiltonians), we estimate $\langle X^m \rangle$ as $L^{3m/2} e /E$.
For $L=500$, we get $\langle X \rangle \approx 1.996$ ($E = 2\times 10^{10}$)
and $\langle X^2 \rangle \approx 5.29$ ($E=4\times 10^{11}$), which should be compared to
the prediction for infinite $L$, $\langle X \rangle = 2$ and $\langle X^2 \rangle
= \frac{16}{3} \approx 5.33$.
Considering that the finite size correction should be $\Or(1/L)$ (see \sref{Sec:FiniteSize}), 
our simulation results are consistent with the predictions.

\section{Derivation of \eref{Eq:TwoPointFin}}
\label{App:TwoPoint}
In this appendix, we calculate the joint probability
$P_2(\genotype_1,\genotype_2)$ that two genotypes $\genotype_1,
\genotype_2$ are both local fitness maxima for FGM with $n=1$.
Let us consider two genotypes with the following sequences,
\begin{eqnarray}
	\genotype_1 &= \{ 
	\stackrel{u_3}{\overbrace{1,1,\cdots, 1}} 
	, \stackrel{u_2}{\overbrace{1,1,\cdots, 1}} 
	, \stackrel{u_1}{\overbrace{0,0,\cdots, 0}} 
	, \stackrel{u_0}{\overbrace{0,0,\cdots, 0}} 	
	\}, \nonumber \\
	\genotype_2 &= \{ 
	\stackrel{u_3}{\overbrace{1,1,\cdots, 1}} 
	, \stackrel{u_2}{\overbrace{0,0,\cdots, 0}} 
	, \stackrel{u_1}{\overbrace{1,1,\cdots, 1}} 
	, \stackrel{u_0}{\overbrace{0,0,\cdots, 0}} 	
	\},
	\label{Eq:align}
\end{eqnarray}
where the $u_i$'s have the same meaning as in \eref{Eq:s0s3}.
We denote the random phenotype variables associated with the sites in
the regions of size $u_0$, $u_1$, $u_2$, and $u_3$ by $\xi_i$, ${\zeta}_j$, ${\eta}_k$, and
${\nu}_l$, respectively. Accordingly, 
the phenotypes ${x}$ and ${y}$ corresponding to the 
genotypes $\genotype_1$ and $\genotype_2$ are
\begin{equation}
	{x} = \sum_{k=1}^{u_2} {\eta}_k + \sum_{l=1}^{u_3} {\nu}_l,
	\quad
	{y} = \sum_{j=1}^{u_1} {\zeta}_j + \sum_{l=1}^{u_3} {\nu}_l.
\end{equation}
Defining
\begin{equation}
	\D({w}, {v})
	= \left \{z \in \R | 
	\left | z - {w} \right | > |{w}| \, \&\,
	\left | z - {v} \right | > |{v}| \right \},
\end{equation} 
$P_2(\genotype_1,\genotype_2)$ can be formally written as
\begin{eqnarray}
\fl
P_2 =& 
\int \rmd  {x} \rmd y 
\prod_{i=1}^{u_0} \int_{\D(-{x},-y)} \rmd \xi_i p(\xi_i) 
\prod_{j=1}^{u_1} \int_{\D(-{x},y)} \rmd \zeta_j p(\zeta_j) 
\prod_{k=1}^{u_2} \int_{\D({x},-y)} \rmd  \eta_k p(\eta_k) 
\nonumber \\
\fl
&\prod_{l=1}^{u_3} \int_{\D({x},y)} \rmd  \nu_l p(\nu_l) 
 \delta\left ({x}-\sum_{k=1}^{u_2} \eta_k-\sum_{l=1}^{u_3}\nu_l \right )
\delta\left (y-\sum_{j=1}^{u_1} \zeta_j-\sum_{i=1}^{u_3}\nu_{l} \right ).
\end{eqnarray}
Using the integral representation of the delta function, we get
\begin{eqnarray}
P_2 =& 
\int \frac{\rmd {k}_x \rmd {k}_y \rmd  {x} \rmd y} {(2\pi)^2}   
	\exp\left ( \rmi k_x  x + \rmi k_y  {y}\right ) \nonumber\\
& I(-{x}, -{y}, 0)^{u_0}
I(-{x}, {y}, {k}_y )^{u_1}
I({x}, -{y}, {k}_x )^{u_2}
I({x}, {y}, {k}_x +k_y)^{u_3},
\label{Eq:TwoPoint}
\end{eqnarray}
where 
\begin{equation}
	I(x, y, k ) = \int_{\D({x},y)} \rmd z p(z) \exp\left (-\rmi kz\right ).
\end{equation}
As discussed in  \ref{App:Moment_Cal}, the dominant contribution for large
$u_i$ comes from 
the region of small $x$, $y$, $k_x$, $k_y$. 
Expanding the integrand for small $x$, $y$, and $k$, we have
\begin{equation}
	I(x, y, k ) = \exp\left (-\frac{ k^2}{2}\right ) - p(0) 
	\epsilon(x, y) \left [ 1  + \Or(k) \right ], 
\end{equation}
where 
\begin{eqnarray}
	\epsilon(x, y) &= \int_{\R \setminus \D(x,y)} \frac{p(z)}{p(0)} d z
	\approx \int_{\R \setminus \D(x,y)} d z\nonumber\\
	&= \max\left (|x|-x,|y|-y \right ) + \max\left (x+|x|,y+|y| \right ).
\end{eqnarray} 
Using these results, the leading contribution to \eref{Eq:TwoPoint} becomes
\begin{eqnarray}
	\fl
P_2 \approx 
\int \frac{\rmd  k_x \rmd  k_y } {2\pi}   
	\exp \left [ -\frac{ u_3 }{2} (k_x + k_y)^2 -\frac{ u_1 }{2} k_x^2-\frac{ u_2 }{2}k_y^2   \right ]
\nonumber\\
	 \int d x d y\exp \left [ - u_0 \epsilon(-x, -y)  - u_1 \epsilon(-x, y) - u_2 \epsilon(x, -y) - u_3 \epsilon(x,y)\right ] \nonumber\\
	 \fl
	= \frac{3}{(u_0+2 u_1+2 u_2+u_3) (2u_0+u_1+u_2+2 u_3)\sqrt{u_1u_2 + u_1 u_3 + u_2 u_3}}.
\end{eqnarray}

Calculating $P_1$ can be done in a similar manner:
\begin{equation}
	P_1(\genotype_1)= 
\int \frac{\rmd {k} \rmd  {x}} {2\pi}   
	\exp\left ( \rmi k  x \right )
	I_1(-x, 0)^{L-d_1} I_2(x, k)^{d_1},
\end{equation}
where $d_1 = u_2+u_3$ and 
\begin{equation}
	I_1(x, k ) = \int_{\D[x]} \rmd z p(z) \exp\left (-\rmi kz\right )
	=\exp\left ( -\frac{k^2}{2} \right ) - 2 p(0) |x| \left [ 1 + \Or(k)\right ].
\end{equation}
For large $d_1$ and $L-d_1$, we obtain
\begin{equation}
	P_1 = \frac{1}{L\sqrt{d_1}}.
\end{equation}
Using $u_0+u_1+u_2+u_3=L$, we arrive at \eref{Eq:TwoPointFin}.

\section*{References}
\bibliographystyle{unsrt}
\bibliography{me}

\begin{thebibliography}{10}

\bibitem{Orr2005}
H.~Allen Orr.
\newblock The genetic theory of adaptation: A brief history.
\newblock {\em Nat. Rev. Genet.}, 6:119--127, 2005.

\bibitem{Visser2014}
J.~A. G.~M. de~Visser and Joachim Krug.
\newblock Empirical fitness landscapes and the predictability of evolution.
\newblock {\em Nat. Rev. Genet.}, 15:480--490, 2014.

\bibitem{Fragata2019}
I.~Fragata, A.~Blanckaert, M.~A.~Dias Louro, D.~A. Liberles, and C.~Bank.
\newblock Evolution in the light of fitness landscape theory.
\newblock {\em Trends Ecol. Evol.}, 34:69--82, 2019.

\bibitem{Wright1931}
S.~Wright.
\newblock Evolution in {M}endelian populations.
\newblock {\em Genetics}, 16:97--159, 1931.

\bibitem{Szendro2013a}
Ivan~G. Szendro, Martijn~F. Schenk, Jasper Franke, Joachim Krug, and
  J.~Arjan~G.M. de~Visser.
\newblock Quantitative analyses of empirical fitness landscapes.
\newblock {\em J. Stat. Mech.:Theory Exp.}, page P01005, 2013.

\bibitem{Gillespie1984}
John~H. Gillespie.
\newblock Molecular evolution over the mutational landscape.
\newblock {\em Evolution}, 38:1116--1129, 1984.

\bibitem{Orr2002}
H.~Allen Orr.
\newblock The population genetics of adaptation: the adaptation of
  \uppercase{DNA} sequences.
\newblock {\em Evolution}, 56:1317--1330, 2002.

\bibitem{Sella2005}
G.~Sella and A.~E. Hirsh.
\newblock The application of statistical physics to evolutionary biology.
\newblock {\em Proc. Nat. Acad. Sci. USA}, 102:9541--9546, 2005.

\bibitem{Stadler1999}
P.~F. Stadler and R.~Happel.
\newblock Random field models for fitness landscapes.
\newblock {\em J. Math. Biol.}, 38:435--478, 1999.

\bibitem{Hwang2018}
S.~Hwang, B.~Schmiegelt, L.~Ferretti, and J.~Krug.
\newblock Universality classes of interaction structures for {NK} fitness
  landscapes.
\newblock {\em J. Stat. Phys.}, 172:226--278, 2018.

\bibitem{Domingo2019}
J.~Domingo, P.~Baeza-Centurion, and B.~Lehner.
\newblock The causes and consequences of genetic interactions (epistasis).
\newblock {\em Ann. Rev. Genom. Hum. Genet.}, 20:17.1--17.28, 2019.

\bibitem{Manrubia2020}
S.~Manrubia, J.A. Cuesta, J.~Aguirre, S.E. Ahnert, L.~Altenberg, A.V. Cano,
  P.~Catal\'{a}n, R.~Diaz-Uriarte, S.F. Elena, J.A. Garc\'{\i}a-Mart\'{\i}n,
  P.~Hogeweg, B.S. Khatri, J.~Krug, A.A. Louis, N.S. Martin, J.L. Payne, M.J.
  Tarnowski, and M.~Wei{\ss}.
\newblock From genotypes to organisms: State-of-the-art and perspectives of a
  cornerstone in evolutionary dynamics.
\newblock {\em Preprint}, arXiv:2002.00363, 2020.

\bibitem{Fisher1930}
R.~A. Fisher.
\newblock {\em The Genetical Theory of Natural Selection}.
\newblock Clarendon Press, Oxford, 1930.

\bibitem{Martin2007}
Guillaume Martin, Santiago~F. Elena, and Thomas Lenormand.
\newblock Distributions of epistasis in microbes fit predictions from a fitness
  landscape model.
\newblock {\em Nat. Gen.}, 39:555--560, 2007.

\bibitem{Gros2009}
Pierre-Alexis Gros, Herv\'{e} {Le Nagard}, and Olivier Tenaillon.
\newblock The evolution of epistasis and its links with genetic robustness,
  complexity and drift in a phenotypic model of adaptation.
\newblock {\em Genetics}, 182:277--293, 2009.

\bibitem{Martin2014}
Guillaume Martin.
\newblock Fisher's geometric model emerges as a property of complex integrated
  phenotypic networks.
\newblock {\em Genetics}, 197:237--255, 2014.

\bibitem{Blanquart2014}
Francois Blanquart, Guillaume Achaz, Thomas Bataillon, and Olivier Tenaillon.
\newblock Properties of selected mutations and genotypic landscapes under
  {F}isher's geometric model.
\newblock {\em Evolution}, 68:3537--3554, 2014.

\bibitem{Tenaillon2014}
Olivier Tenaillon.
\newblock The utility of {F}isher's geometric model in evolutionary genetics.
\newblock {\em Annu. Rev. Ecol. Evol. Syst.}, 45:179--201, 2014.

\bibitem{Hwang2017}
Sungmin Hwang, Su-Chan Park, and Joachim Krug.
\newblock Genotypic complexity of {F}isher{\textquoteright}s geometric model.
\newblock {\em Genetics}, 206(2):1049–1079, 2017.

\bibitem{Schoustra2016}
Sijmen Schoustra, Sungmin Hwang, Joachim Krug, and J.~Arjan~G.M. de~Visser.
\newblock Diminishing-returns epistasis among random beneficial mutations in a
  multicellular fungus.
\newblock {\em Proc. R. Soc. Lond. Ser. B}, 283:20161376, 2016.

\bibitem{Hopfield1982}
J.~J. Hopfield.
\newblock Neural networks and physical systems with emergent collective
  computational abilities.
\newblock {\em Proc. Nat. Acad. Sci. USA}, 79(8):2554--2558, April 1982.

\bibitem{Amit1989}
Daniel~J. Amit.
\newblock {\em Modeling brain function: The world of attractor neural
  networks}.
\newblock Cambridge University Press, Cambridge, UK, 1989.

\bibitem{Hertz1991}
J.~A. Hertz, A.~Krogh, and R.~G. Palmer.
\newblock {\em Introduction to the theory of neural computation}.
\newblock Taylor-Francis, Boca Raton, 1991.

\bibitem{Nokura1998}
Kazuo Nokura.
\newblock {Spin glass states of the anti-Hopfield model}.
\newblock {\em J. Phys. A}, 31:7447--7459, 1998.

\bibitem{Cherrier2003}
R.~Cherrier, D.~S. Dean, and A.~Lef\`evre.
\newblock The number of metastable states in the generalized random orthogonal
  model.
\newblock {\em J. Phys. A: Math. Gen.}, 36(14):3935, 2003.

\bibitem{Challet1997}
D.~Challet and Y.~C. Zhang.
\newblock Emergence of cooperation and organization in an evolutionary game.
\newblock {\em Physica A}, 246(3):407--418, December 1997.

\bibitem{Marsili2000}
Matteo Marsili, Damien Challet, and Riccardo Zecchina.
\newblock Exact solution of a modified {El} {Farol}'s bar problem: {Efficiency}
  and the role of market impact.
\newblock {\em Physica A}, 280(34):522--553, June 2000.

\bibitem{Challet2000}
Damien Challet, Matteo Marsili, and Riccardo Zecchina.
\newblock Statistical {Mechanics} of {Systems} with {Heterogeneous} {Agents}:
  {Minority} {Games}.
\newblock {\em Phys. Rev. Lett.}, 84(8):1824--1827, February 2000.

\bibitem{Chakraborti2015}
Anirban Chakraborti, Damien Challet, Arnab Chatterjee, Matteo Marsili, Yi-Cheng
  Zhang, and Bikas~K. Chakrabarti.
\newblock Statistical mechanics of competitive resource allocation using
  agent-based models.
\newblock {\em Physics Reports}, 552:1 -- 25, 2015.

\bibitem{Ferreira1998}
F.~F. Ferreira and J.~F. Fontanari.
\newblock Probabilistic analysis of the number partitioning problem.
\newblock {\em J. Phys. A}, 31:3417--3428, 1998.

\bibitem{Mertens2000}
S.~Mertens.
\newblock Random costs in combinatorial optimization.
\newblock {\em Phys. Rev. Lett.}, 84:1347--1350, 2000.

\bibitem{Tanaka1980}
F~Tanaka and S~F Edwards.
\newblock Analytic theory of the ground state properties of a spin glass. {I}.
  {I}sing spin glass.
\newblock {\em J. Phys. F: Met. Phys.}, 10(12):2769, 1980.

\bibitem{Bray1980}
A~J Bray and M~A Moore.
\newblock Metastable states in spin glasses.
\newblock {\em J. Phys. C: Solid State Phys.}, 13(19):L469--L476, 1980.

\bibitem{Bray1981}
A.~J. Bray and M.~A. Moore.
\newblock Metastable states in spin glasses with short-ranged interactions.
\newblock {\em J. Phys. C: Solid State Phys.}, 14(9):1313, 1981.

\bibitem{Gardner1986}
E.~Gardner.
\newblock Structure of metastable states in the {Hopfield} model.
\newblock {\em J. Phys. A: Math. Gen.}, 19(16):L1047, 1986.

\bibitem{Treves1988}
A.~Treves and D.~J. Amit.
\newblock Metastable states in asymmetrically diluted {Hopfield} networks.
\newblock {\em J. Phys. A: Math. Gen.}, 21(14):3155, 1988.

\bibitem{Singh1995}
Manoranjan~P. Singh, Zhang Chengxiang, and Chandan Dasgupta.
\newblock Fixed points in a {Hopfield} model with random asymmetric
  interactions.
\newblock {\em Phys. Rev. E}, 52(5):5261--5272, November 1995.

\bibitem{Blanquart2016}
Francois Blanquart and Thomas Bataillon.
\newblock Epistasis and the structure of fitness landscapes: are experimental
  fitness landscapes compatible with {F}isher's geometric model?
\newblock {\em Genetics}, 203:847--862, 2016.

\bibitem{Weinreich2013}
Daniel~M. Weinreich and Jennifer~L. Knies.
\newblock Fisher's geometric model of adaptation meets the functional
  synthesis: Data on pairwise epistasis for fitness yields insights into the
  shape and size of phenotype space.
\newblock {\em Evolution}, 67:2957--2972, 2013.

\bibitem{Erdelyi1955}
A~Erd\'elyi, editor.
\newblock {\em Higher Transcendental Functions}, volume~1.
\newblock McGraw-Hill, New York, 1955.

\bibitem{Nowak2015}
Stefan Nowak and Joachim Krug.
\newblock Analysis of adaptive walks on {NK} fitness landscapes with different
  interaction schemes.
\newblock {\em J. Stat. Mech.:Theory Exp.}, 2015:P06014, 2015.

\bibitem{Kingman1978}
J.~F.~C. Kingman.
\newblock A simple model for the balance between selection and mutation.
\newblock {\em J. Appl. Prob.}, 15:1--12, 1978.

\bibitem{Kauffman1987}
S.~Kauffman and S.~Levin.
\newblock Towards a general theory of adaptive walks on rugged landscapes.
\newblock {\em J. Theor. Biol.}, 128:11--45, 1987.

\bibitem{Derrida1981}
B.~Derrida.
\newblock Random-energy model - an exactly solvable model of disordered
  systems.
\newblock {\em Phys. Rev. B}, 24(5):2613--2626, 1981.

\bibitem{Macken1989}
C.~A. Macken and A.~S. Perelson.
\newblock Protein evolution on rugged landscapes.
\newblock {\em Proc. Nat. Acad. Sci. USA}, 86:6191--6195, 1989.

\bibitem{Baldi1989}
P.~Baldi and Y.~Rinott.
\newblock Asymptotic normality of some graph-related statistics.
\newblock {\em J. Appl. Prob.}, 26:171--175, 1989.

\bibitem{Perelson1995}
A.~S. Perelson and C.~A. Macken.
\newblock Protein evolution on partially correlated landscapes.
\newblock {\em Proc. Nat. Acad. Sci. USA}, 92:9657--9661, 1995.

\bibitem{Schmiegelt2014}
B.~Schmiegelt and J.~Krug.
\newblock Evolutionary accessibility of modular fitness landscapes.
\newblock {\em J. Stat. Phys.}, 154:334--355, 2014.

\bibitem{HwangUn}
S~Hwang, D.~S. Dean, and J~Krug.
\newblock (unpublished).

\bibitem{Zagorski2016}
Marcin Zagorski, Zdzislaw Burda, and Bartlomiej Waclaw.
\newblock Beyond the hypercube: Evolutionary accessibility of fitness
  landscapes with realistic mutational networks.
\newblock {\em PLoS Comp. Biol.}, 12:e1005218, 2016.

\bibitem{Schmiegelt2019}
B.~Schmiegelt and J.~Krug.
\newblock Accessibility percolation on cartesian power graphs.
\newblock {\em Preprint}, arXiv:1912.07925, 2020.

\bibitem{HwangUn2}
S~Hwang and J~Krug.
\newblock (unpublished).

\bibitem{Pokusaeva2019}
Victoria~O. Pokusaeva, Dinara~R. Usmanova, Ekaterina~V. Putintseva, Lorena
  Espinar, Karen~S. Sarkisyan, Alexander~S. Mishin, Natalya~S. Bogatyreva,
  Dmitry~N. Ivankov, Arseniy~V. Akopyan, Sergey~Ya. Avvakumov, Inna~S.
  Povolotskaya, Guillaume~J. Filion, Lucas~B. Carey, and Fyodor~A. Kondrashov.
\newblock An experimental assay of the interactions of amino acids from
  orthologous sequences shaping a complex fitness landscape.
\newblock {\em PLoS Genet.}, 15:e1008079, 2019.

\end{thebibliography}
\end{document}